\pdfoutput=1
\documentclass[12pt,a4paper]{article}

\usepackage{xcolor}
\usepackage{nicefrac}
\usepackage{afterpage}
\usepackage{rotating}
\usepackage{siunitx}
\newcommand{\round}[2]{\num[round-mode=places,round-precision=#1]{#2}}

\usepackage{ifthen}
\newboolean{pdflatex}
\setboolean{pdflatex}{true}

\newboolean{articletitles}
\setboolean{articletitles}{true}

\newboolean{uprightparticles}
\setboolean{uprightparticles}{false}

\def\paperauthors{LHCb collaboration}
\def\paperasciititle{Measurement of the relative branching fractions of B+ -> h+ h'+ h'-$ decays}
\def\papertitle{Measurement of the relative branching fractions of~$\Bp \to h^+h^{\prime +}h^{\prime -}$~decays}
\def\paperkeywords{{High Energy Physics}, {LHCb}}
\def\papercopyright{\the\year\ CERN for the benefit of the LHCb collaboration}
\def\paperlicence{CC BY 4.0 licence}
\def\paperlicenceurl{https://creativecommons.org/licenses/by/4.0/}

\usepackage[top=1in, bottom=1.25in, left=1in, right=1in]{geometry}

\usepackage{microtype}
\usepackage{lineno}  
\usepackage{xspace} 
\usepackage{caption} 

\usepackage{graphicx}  
\usepackage{color}
\usepackage{colortbl}
\graphicspath{{./figs/}} 

\usepackage{amsmath} 
\usepackage{amssymb}
\usepackage{amsfonts}
\usepackage{upgreek} 

\newcommand*\patchAmsMathEnvironmentForLineno[1]{%
\expandafter\let\csname old#1\expandafter\endcsname\csname #1\endcsname
\expandafter\let\csname oldend#1\expandafter\endcsname\csname
end#1\endcsname
 \renewenvironment{#1}%
   {\linenomath\csname old#1\endcsname}%
   {\csname oldend#1\endcsname\endlinenomath}%
}
\newcommand*\patchBothAmsMathEnvironmentsForLineno[1]{%
  \patchAmsMathEnvironmentForLineno{#1}%
  \patchAmsMathEnvironmentForLineno{#1*}%
}
\AtBeginDocument{%
\patchBothAmsMathEnvironmentsForLineno{equation}%
\patchBothAmsMathEnvironmentsForLineno{align}%
\patchBothAmsMathEnvironmentsForLineno{flalign}%
\patchBothAmsMathEnvironmentsForLineno{alignat}%
\patchBothAmsMathEnvironmentsForLineno{gather}%
\patchBothAmsMathEnvironmentsForLineno{multline}%
\patchBothAmsMathEnvironmentsForLineno{eqnarray}%
}

\usepackage{hyperxmp}

\usepackage[pdftex,
            pdfauthor={\paperauthors},
            pdftitle={\paperasciititle},
            pdfkeywords={\paperkeywords},
            pdfcopyright={Copyright (C) \papercopyright},
            pdflicenseurl={\paperlicenceurl}]{hyperref}

\usepackage[colorinlistoftodos,textsize=scriptsize]{todonotes}

\usepackage[bottom,flushmargin,hang,multiple]{footmisc}

\usepackage[all]{hypcap} 

\usepackage{xspace} 
\usepackage{upgreek}

\def\lhcb   {\mbox{LHCb}\xspace}

\def\MagUp {\mbox{\em Mag\kern -0.05em Up}\xspace}

\ifthenelse{\boolean{uprightparticles}}%
{

 \def\Peta        {\ensuremath{\upeta}\xspace}

 \def\Ppi         {\ensuremath{\uppi}\xspace}

 \def\Ppsi        {\ensuremath{\uppsi}\xspace}

 \def\PDelta      {\ensuremath{\Delta}\xspace}                 
 \def\PXi         {\ensuremath{\Xi}\xspace}                 
 \def\PLambda     {\ensuremath{\Lambda}\xspace}                 
 \def\PSigma      {\ensuremath{\Sigma}\xspace}                 
 \def\POmega      {\ensuremath{\Omega}\xspace}                 
 \def\PUpsilon    {\ensuremath{\Upsilon}\xspace}

 \def\PB      {\ensuremath{\mathrm{B}}\xspace}                 
                  
 \def\PD      {\ensuremath{\mathrm{D}}\xspace}

 \def\PJ      {\ensuremath{\mathrm{J}}\xspace}                 
 \def\PK      {\ensuremath{\mathrm{K}}\xspace}

 \def\Pb      {\ensuremath{\mathrm{b}}\xspace}                 
 \def\Pc      {\ensuremath{\mathrm{c}}\xspace}

 \def\Ph      {\ensuremath{\mathrm{h}}\xspace}                 
 \def\Pi      {\ensuremath{\mathrm{i}}\xspace}

 \def\Pp      {\ensuremath{\mathrm{p}}\xspace}

 \def\Ps      {\ensuremath{\mathrm{s}}\xspace}

 \def\thebaroffset{0.0em}
}
{

 \def\Peta        {\ensuremath{\eta}\xspace}

 \def\Ppi         {\ensuremath{\pi}\xspace}

 \def\Ppsi        {\ensuremath{\psi}\xspace}                 
                  
 \mathchardef\PDelta="7101
 \mathchardef\PXi="7104
 \mathchardef\PLambda="7103
 \mathchardef\PSigma="7106
 \mathchardef\POmega="710A
 \mathchardef\PUpsilon="7107
                  
 \def\PB      {\ensuremath{B}\xspace}                 
                  
 \def\PD      {\ensuremath{D}\xspace}

 \def\PJ      {\ensuremath{J}\xspace}                 
 \def\PK      {\ensuremath{K}\xspace}

 \def\Pb      {\ensuremath{b}\xspace}                 
 \def\Pc      {\ensuremath{c}\xspace}

 \def\Ph      {\ensuremath{h}\xspace}                 
 \def\Pi      {\ensuremath{i}\xspace}

 \def\Pp      {\ensuremath{p}\xspace}

 \def\Ps      {\ensuremath{s}\xspace}

 \def\thebaroffset{0.18em}
}
\newcommand{\offsetoverline}[2][\thebaroffset]{\kern #1\overline{\kern -#1 #2}}%

\makeatletter
\ifcase \@ptsize \relax
  \newcommand{\miniscule}{\@setfontsize\miniscule{4}{5}}
\or
  \newcommand{\miniscule}{\@setfontsize\miniscule{5}{6}}
\or
  \newcommand{\miniscule}{\@setfontsize\miniscule{5}{6}}
\fi
\makeatother

\DeclareRobustCommand{\optbar}[1]{\shortstack{{\miniscule (\rule[.5ex]{1.25em}{.18mm})}
  \\ [-.7ex] $#1$}}

\def\squark    {{\ensuremath{\Ps}}\xspace}

\def\cquark    {{\ensuremath{\Pc}}\xspace}

\def\bquark    {{\ensuremath{\Pb}}\xspace}

\def\hadron {{\ensuremath{\Ph}}\xspace}
\def\pion   {{\ensuremath{\Ppi}}\xspace}

\def\pip    {{\ensuremath{\pion^+}}\xspace}
\def\pim    {{\ensuremath{\pion^-}}\xspace}

\def\kaon    {{\ensuremath{\PK}}\xspace}
\def\Kbar    {{\ensuremath{\offsetoverline{\PK}}}\xspace}

\def\KorKbar {\kern \thebaroffset\optbar{\kern -\thebaroffset \PK}{}\xspace}

\def\Kp      {{\ensuremath{\kaon^+}}\xspace}
\def\Km      {{\ensuremath{\kaon^-}}\xspace}

\def\Kstar   {{\ensuremath{\kaon^*}}\xspace}
\def\Kstarb  {{\ensuremath{\Kbar{}^*}}\xspace}

\newcommand{\etapr}{\ensuremath{\Peta^{\prime}}\xspace}

\def\Dbar    {{\ensuremath{\offsetoverline{\PD}}}\xspace}
\def\D       {{\ensuremath{\PD}}\xspace}

\def\DorDbar {\kern \thebaroffset\optbar{\kern -\thebaroffset \PD}\xspace}
\def\Dz      {{\ensuremath{\D^0}}\xspace}
\def\Dzb     {{\ensuremath{\Dbar{}^0}}\xspace}
\def\Dp      {{\ensuremath{\D^+}}\xspace}
\def\Dm      {{\ensuremath{\D^-}}\xspace}

\def\DpDm    {\ensuremath{\Dp {\kern -0.16em \Dm}}\xspace}

\def\Dstarp  {{\ensuremath{\D^{*+}}}\xspace}

\def\Dsm     {{\ensuremath{\D^-_\squark}}\xspace}

\def\B       {{\ensuremath{\PB}}\xspace}

\def\BorBbar {\kern \thebaroffset\optbar{\kern -\thebaroffset \PB}\xspace}
\def\Bz      {{\ensuremath{\B^0}}\xspace}

\def\Bd      {{\ensuremath{\B^0}}\xspace}

\def\BdorBdbar {\kern \thebaroffset\optbar{\kern -\thebaroffset \Bd}\xspace}
\def\Bu      {{\ensuremath{\B^+}}\xspace}

\def\Bp      {{\ensuremath{\Bu}}\xspace}

\def\Bs      {{\ensuremath{\B^0_\squark}}\xspace}

\def\BsorBsbar {\kern \thebaroffset\optbar{\kern -\thebaroffset \Bs}\xspace}

\def\jpsi     {{\ensuremath{{\PJ\mskip -3mu/\mskip -2mu\Ppsi}}}\xspace}

\def\Y#1S{\ensuremath{\PUpsilon{(#1S)}}\xspace}

\def\proton      {{\ensuremath{\Pp}}\xspace}
\def\antiproton  {{\ensuremath{\overline \proton}}\xspace}

\def\LorLbar     {\kern \thebaroffset\optbar{\kern -\thebaroffset \PLambda}\xspace}

\def\Xires       {{\ensuremath{\PXi}}\xspace}

\def\Xiresbar       {{\ensuremath{\offsetoverline{\Xires}}}\xspace}

\def\Xibbarp      {{\ensuremath{\Xiresbar{}_\bquark^+}}\xspace}

\def\BF         {{\ensuremath{\mathcal{B}}}\xspace}

\newcommand{\decay}[2]{\ensuremath{#1\!\to #2}\xspace} 

\def\to                 {\ensuremath{\rightarrow}\xspace}

\def\CP                {{\ensuremath{C\!P}}\xspace}

\def\AT#1     {\ensuremath{A_{\mathrm{T}}^{#1}}\xspace}           

\def\C#1      {\ensuremath{\mathcal{C}_{#1}}\xspace}                       
\def\Cp#1     {\ensuremath{\mathcal{C}_{#1}^{'}}\xspace}                    
\def\Ceff#1   {\ensuremath{\mathcal{C}_{#1}^{\mathrm{(eff)}}}\xspace}        
\def\Cpeff#1  {\ensuremath{\mathcal{C}_{#1}^{'\mathrm{(eff)}}}\xspace}       
\def\Ope#1    {\ensuremath{\mathcal{O}_{#1}}\xspace}                       
\def\Opep#1   {\ensuremath{\mathcal{O}_{#1}^{'}}\xspace}                    

\newcommand{\nospaceunit}[1]{\ensuremath{\text{#1}}}       
\newcommand{\aunit}[1]{\ensuremath{\text{\,#1}}}       

\newcommand{\tev}{\aunit{Te\kern -0.1em V}\xspace}
\newcommand{\gev}{\aunit{Ge\kern -0.1em V}\xspace}
\newcommand{\mev}{\aunit{Me\kern -0.1em V}\xspace}
\newcommand{\kev}{\aunit{ke\kern -0.1em V}\xspace}
\newcommand{\ev}{\aunit{e\kern -0.1em V}\xspace}
 
\newcommand{\mevc}{\ensuremath{\aunit{Me\kern -0.1em V\!/}c}\xspace}
\newcommand{\gevc}{\ensuremath{\aunit{Ge\kern -0.1em V\!/}c}\xspace}
\newcommand{\mevcc}{\ensuremath{\aunit{Me\kern -0.1em V\!/}c^2}\xspace}
\newcommand{\gevcc}{\ensuremath{\aunit{Ge\kern -0.1em V\!/}c^2}\xspace}

\def\mum  {\ensuremath{\,\upmu\nospaceunit{m}}\xspace}

\def\fb   {\ensuremath{\aunit{fb}}\xspace}
\def\invfb   {\ensuremath{\fb^{-1}}\xspace}

\newcommand{\stat}{\aunit{(stat)}\xspace}
\newcommand{\syst}{\aunit{(syst)}\xspace}

\newcommand{\chisq}{\ensuremath{\chi^2}\xspace}

\newcommand{\chisqip}{\ensuremath{\chi^2_{\text{IP}}}\xspace}

\def\gsim{{~\raise.15em\hbox{$>$}\kern-.85em
          \lower.35em\hbox{$\sim$}~}\xspace}
\def\lsim{{~\raise.15em\hbox{$<$}\kern-.85em
          \lower.35em\hbox{$\sim$}~}\xspace}

\def\sWeights{\mbox{\em sWeights}\xspace}

\def\pt         {\ensuremath{p_{\mathrm{T}}}\xspace}

\def\ptot       {\ensuremath{p}\xspace}

\def\evtgen     {\mbox{\textsc{EvtGen}}\xspace}

\def\geant      {\mbox{\textsc{Geant4}}\xspace}

\def\neurobayes {\mbox{\textsc{NeuroBayes}}\xspace}
\def\photos     {\mbox{\textsc{Photos}}\xspace}

\def\pythia     {\mbox{\textsc{Pythia}}\xspace}

\def\tell1  {TELL1\xspace}
\def\ukl1   {UKL1\xspace}

\newcommand{\ie}{\mbox{\itshape i.e.}\xspace}

\def\phntm      {\phantom{0}}
\def\phntn      {\phantom{$-$}}

\def\hhh 	  	{\hadron^+\hadron^{\prime +}\hadron^{\prime -}}

\def\BuTohhh 	  	{\decay{\Bp}{\hhh}}
\def\BuToKKK 	  	{\decay{\Bp}{\Kp \Kp \Km}} 
\def\BuToKpiK 	  	{\decay{\Bp}{\pip \Kp \Km}} 
\def\BuToKpipi 	  	{\decay{\Bp}{\Kp \pip \pim}} 
\def\BuTopipipi 	{\decay{\Bp}{\pip \pip \pim}}

\def\mpr                {\ensuremath{m^\prime}}
\def\thetapr            {\ensuremath{\theta^\prime}}

\usepackage{cite} 
\usepackage{mciteplus}

\begin{document}

\renewcommand{\thefootnote}{\fnsymbol{footnote}}
\setcounter{footnote}{1}
\begin{titlepage}
\pagenumbering{roman}

\vspace*{-1.5cm}
\centerline{\large EUROPEAN ORGANIZATION FOR NUCLEAR RESEARCH (CERN)}
\vspace*{1.5cm}
\noindent
\begin{tabular*}{\linewidth}{lc@{\extracolsep{\fill}}r@{\extracolsep{0pt}}}
\ifthenelse{\boolean{pdflatex}}
{\vspace*{-1.5cm}\mbox{\!\!\!\includegraphics[width=.14\textwidth]{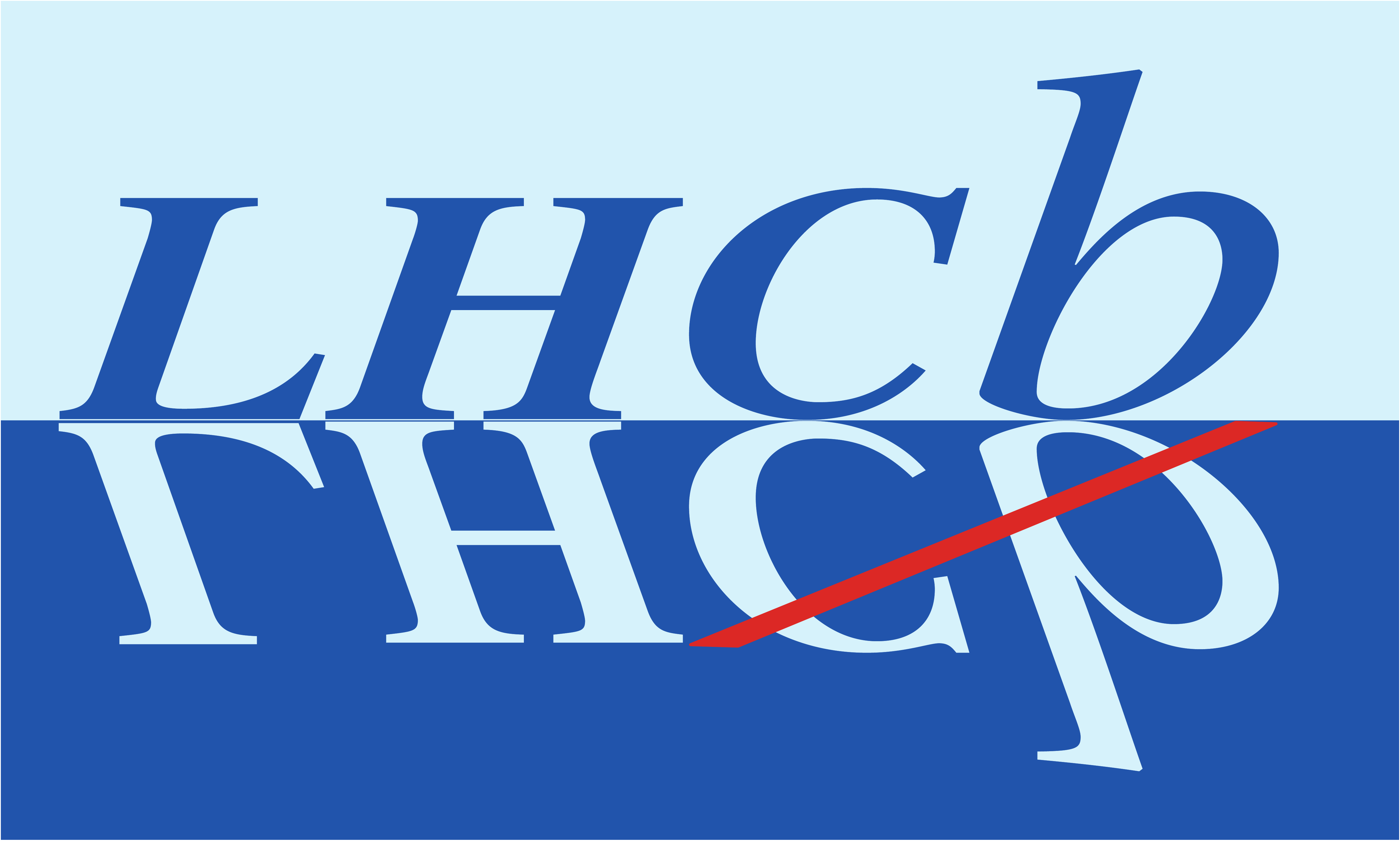}} & &}%
{\vspace*{-1.2cm}\mbox{\!\!\!\includegraphics[width=.12\textwidth]{figs/lhcb-logo.eps}} & &}%
\\
 & & CERN-EP-2020-189 \\  
 & & LHCb-PAPER-2020-031 \\  
 & & October 23, 2020 \\ 
 & & \\
\end{tabular*}

\vspace*{4.0cm}

{\normalfont\bfseries\boldmath\huge
\begin{center}
  \papertitle 
\end{center}
}

\vspace*{2.0cm}

\begin{center}
\paperauthors\footnote{Authors are listed at the end of this paper.}
\end{center}

\vspace{\fill}

\begin{abstract}
  \noindent
  The relative branching fractions of $\Bp \to h^+h^{\prime +}h^{\prime -}$ decays, where $h^{(\prime)}$ is a pion or kaon, are measured.
  The analysis is performed with a data sample, collected with the LHCb detector, corresponding to an integrated luminosity of $3.0 \invfb$ of $pp$ collisions.
  The results obtained improve significantly on previous measurements of these quantities, and are important for the interpretation of Dalitz plot analyses of three-body charmless hadronic decays of \Bp\ mesons.
\end{abstract}

\vspace*{2.0cm}

\begin{center}
  Published in Phys.\ Rev.\ D102 (2020) 112010
\end{center}

\vspace{\fill}

{\footnotesize 
\centerline{\copyright~\papercopyright. \href{\paperlicenceurl}{\paperlicence}.}}
\vspace*{2mm}

\end{titlepage}

\newpage
\setcounter{page}{2}
\mbox{~}

\renewcommand{\thefootnote}{\arabic{footnote}}
\setcounter{footnote}{0}
\cleardoublepage


\pagestyle{plain} 
\setcounter{page}{1}
\pagenumbering{arabic}

\section{Introduction}
\label{sec:Introduction}

Three-body hadronic \B\ meson decays to final states without any charm or charmonium hadrons are of great interest since they can be mediated by both tree and loop (so-called penguin) diagrams, and consequently \CP-violation effects can manifest.
Such charmless three-body decays can proceed through a number of different intermediate resonances, which increases the range of ways in which \CP-violation effects can occur.
Model-independent studies of the $\Bp \to \Kp\Kp\Km$, $\pip\Kp\Km$, $\Kp\pip\pim$ and $\pip\pip\pim$ decays, collectively referred to as $\Bp \to h^+h^{\prime +}h^{\prime -}$ decays, have revealed large \CP-violation effects in certain regions of their Dalitz plots~\cite{LHCb-PAPER-2013-027,LHCb-PAPER-2013-051,LHCb-PAPER-2014-044}, with these results confirmed for $\Bp \to \pip\Kp\Km$ and $\pip\pip\pim$ decays by model-dependent Dalitz-plot analyses~\cite{LHCb-PAPER-2018-051,LHCb-PAPER-2019-017,LHCb-PAPER-2019-018}.\footnote{The inclusion of charge-conjugate processes is implied throughout this paper.}
It is as yet unclear whether the observed effects can be explained within the Standard Model or if new dynamics are involved.

The results of Dalitz-plot analyses typically include fit fractions of contributing resonances.
These can be converted to quasi-two-body branching fractions, which can be predicted theoretically (see, for example, Refs.~\cite{Beneke:2003zv,Chiang:2008zb,Cheng:2010yd,Ebert:2011ry,Zou:2012td,Cheng:2013dua,Cheng:2016shb,Wang:2016rlo}), by multiplication with the branching fraction for the three-body decay.
Interpretation of the data requires both branching fractions and \CP\ asymmetries to be considered.
Consequently, precise measurements of the branching fractions of charmless hadronic three-body \Bp\ decays are needed.
Current knowledge of the $\Bp \to h^+h^{\prime +}h^{\prime -}$ branching fractions, as tabulated by the Particle Data Group (PDG)~\cite{PDG2020}, is summarised in Table~\ref{tab:PDG-BFs}.
The precision ranges from 4\,\% to 9\,\%, which is not sufficient given the sensitivity of the most recent Dalitz-plot analyses.
Improved knowledge of these quantities is therefore required.

\begin{table}[!b]
\centering
\caption{
  Current knowledge of the branching fractions of $\Bp \to h^+h^{\prime +}h^{\prime -}$ decays~\cite{PDG2020}.
}
\label{tab:PDG-BFs}
\begin{tabular}{lcc}
  \hline
  \hline
  Decay & PDG average $(10^{-6})$ & References \\
  \hline
  $\Bp \to \Kp\Kp\Km$    & $34.0 \pm 1.4$     & \cite{Garmash:2004wa,Lees:2012kxa} \\
  $\Bp \to \pip\Kp\Km$   & \phntm$5.2 \pm0.4$ & \cite{Aubert:2007xb,Hsu:2017kir} \\
  $\Bp \to \Kp\pip\pim$  & $51.0 \pm 2.9$     & \cite{Garmash:2005rv,Aubert:2008bj} \\
  $\Bp \to \pip\pip\pim$ & $15.2 \pm 1.4$     & \cite{Aubert:2009av} \\
  \hline
  \hline
\end{tabular}
\end{table}

\begin{figure}[!tb]
  \centering
  \includegraphics[width=0.48\textwidth]{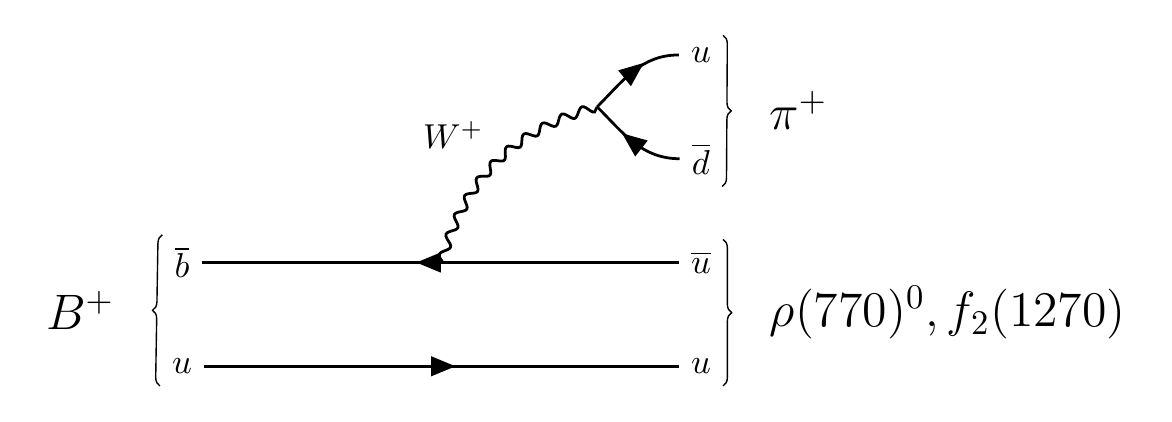}
  \includegraphics[width=0.48\textwidth]{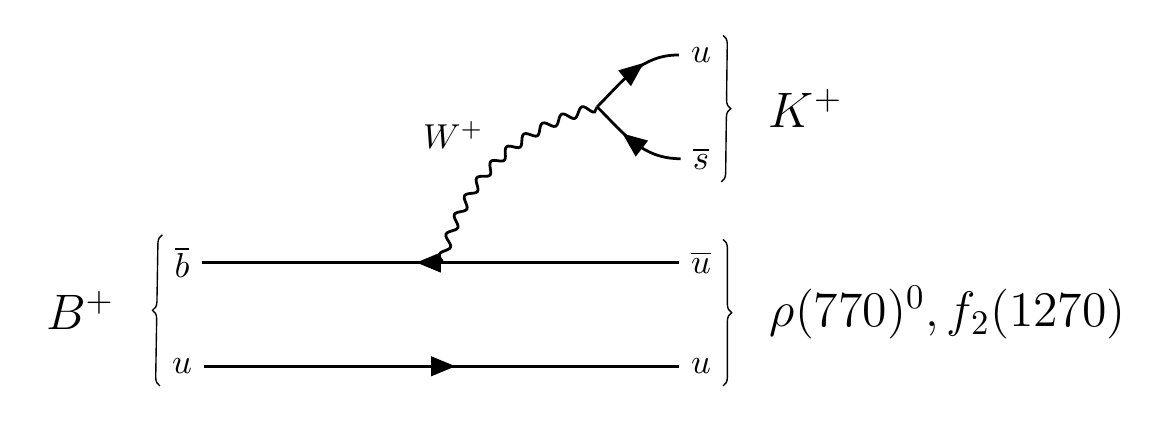} \\

  \includegraphics[width=0.48\textwidth]{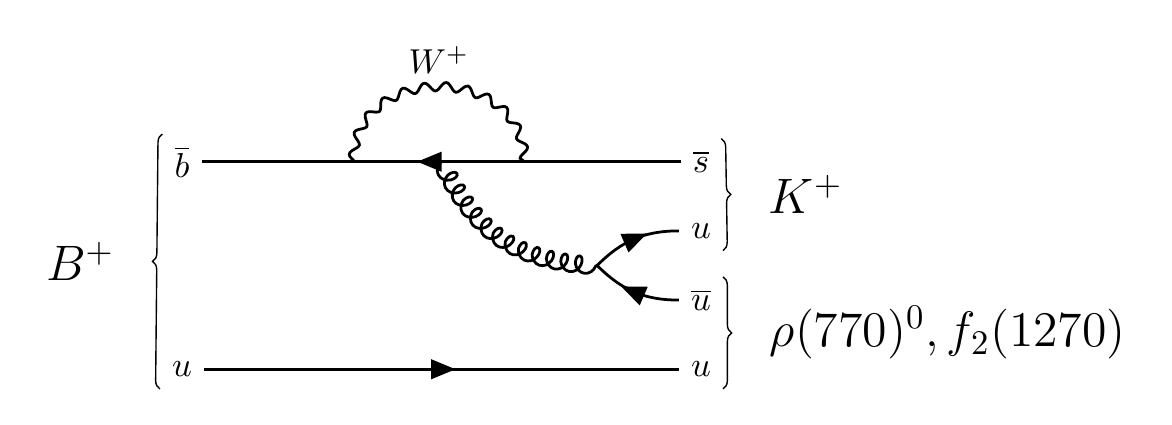}
  \includegraphics[width=0.48\textwidth]{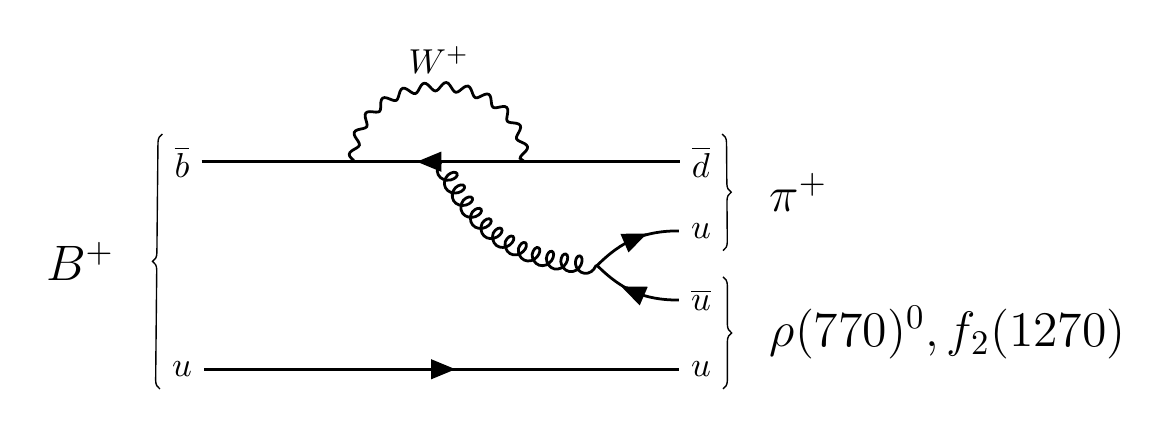} \\

  \includegraphics[width=0.48\textwidth]{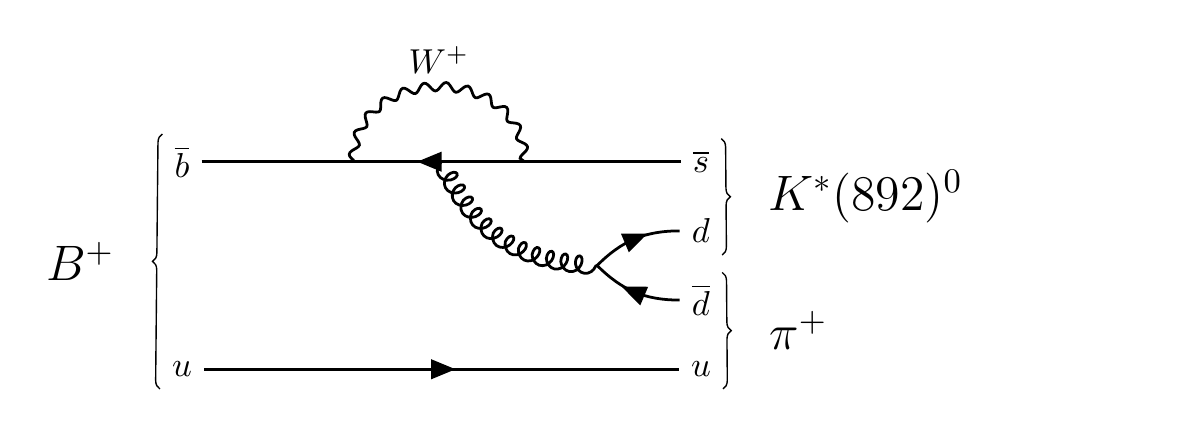}
  \includegraphics[width=0.48\textwidth]{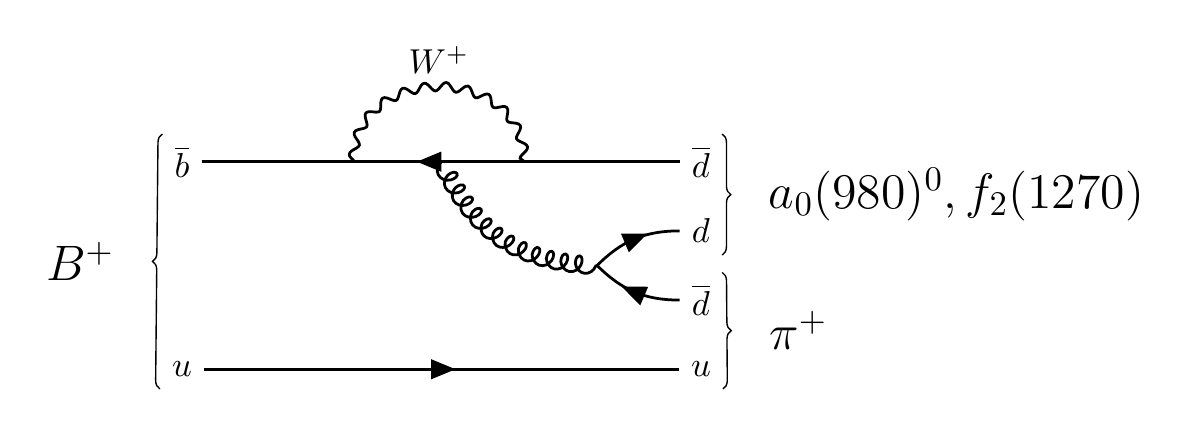} \\

  \includegraphics[width=0.48\textwidth]{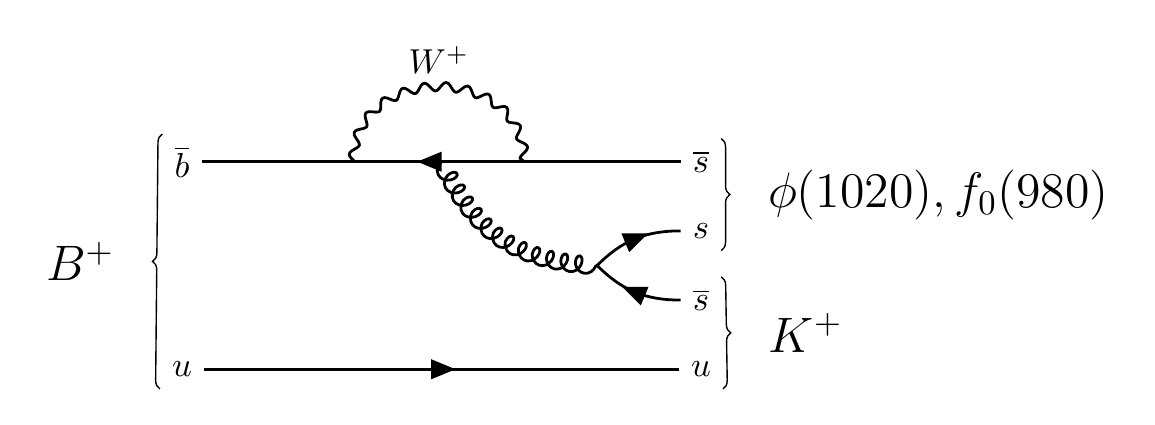}
  \includegraphics[width=0.48\textwidth]{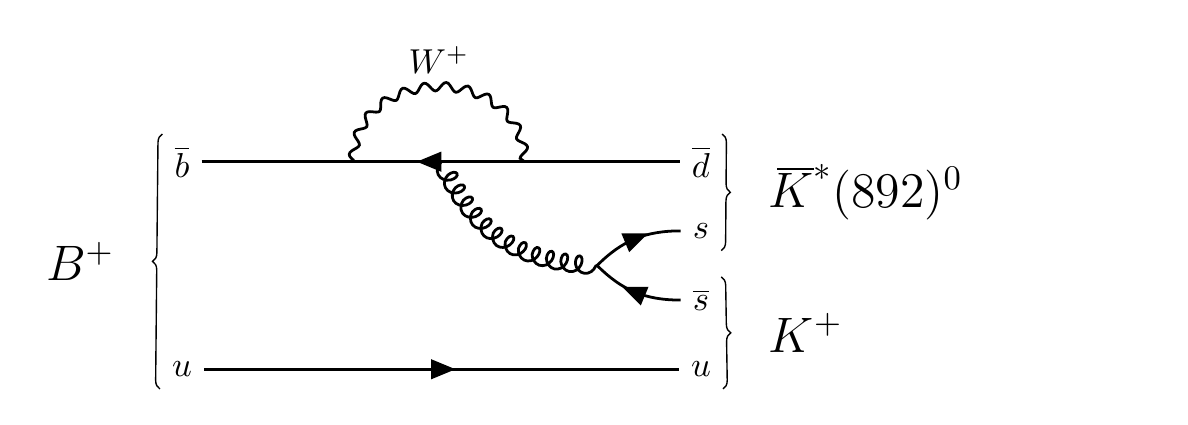}

  \caption{
    Example Feynman diagrams that contribute to $\Bp \to h^+h^{\prime +}h^{\prime -}$ decays.
    (Top row)~tree-level processes with external $W$ emission coupling to (left) pion and (right) kaon;
    (second row) (left)~$\bar{b} \to \bar{s}$ and (right) $\bar{b} \to \bar{d}$ loop-level processes with $u\bar{u}$ production;
    (following rows)~same, but with (third row)~$d\bar{d}$ and (bottom row)~$s\bar{s}$ production.
    Where final-state particles other than $\pip$ and $\Kp$ are given, it should be understood that a range of resonances are possible, and where these are unflavoured in many cases decays to both $\pip\pim$ and $\Kp\Km$ are possible.
    Other types of Feynman diagrams that can also contribute, such as internal $W$ emission and annihilation amplitudes as well as rescattering processes, are not shown.
  }
  \label{fig:feynman}
\end{figure}

The relative size of the branching fractions of $\Bp \to h^+h^{\prime +}h^{\prime -}$ decays, as given in Table~\ref{tab:PDG-BFs}, can be understood to first approximation through the Cabibbo--Kobayashi--Maskawa matrix elements that enter the relevant Feynman diagrams.
Examples of such diagrams are shown in Fig.~\ref{fig:feynman}.
Interference between different amplitudes contributing to the same process can cause \CP\ violation.

In this paper, the relative branching fractions of the \mbox{$\Bp \to h^+h^{\prime +}h^{\prime -}$} decays are determined.
The analysis is based on a data sample corresponding to an integrated luminosity of $3.0\invfb$ of $pp$ collision data collected with the LHCb detector, of which $1.0 \invfb$ was collected in 2011 when the centre-of-mass energy, $\sqrt{s}$, was equal to $7\tev$ and the remaining $2.0 \invfb$ was collected in 2012 at $\sqrt{s} = 8 \tev$.
Since currently \mbox{${\cal B}(\Bp \to \Kp\Kp\Km)$} is known most precisely, results are presented primarily as ratios with this mode as the denominator.
However, determinations of all ratios of one mode to another are also presented, as are the correlations between the results, in order to profit from future improvements of any of the individual branching fraction measurements.
The analysis presented here does not include study of the suppressed three-body charmless hadronic decays $\Bp \to \Kp\Kp\pim$ and $\Bp \to \pip\pip\Km$, which require dedicated measurements~\cite{Garmash:2003er,Aubert:2008rr,LHCb-PAPER-2016-023}.

Previous measurements have used slightly different definitions of the three-body branching fractions, ${\cal B}(\Bp \to h^+h^{\prime +}h^{\prime -})$, and given the desired precision it is important to have a clear definition.
In the work presented here, any $\Bp \to h^+h^{\prime +}h^{\prime -}$ decay where the three final-state particles originate from the same vertex is considered to be part of the signal.
This definition thus includes all charmonium resonances, since all have negligible lifetimes, and excludes all contributions from weakly decaying charm mesons.
This choice differs from that used in some Dalitz-plot analyses, where contributions from the \jpsi\ resonance are often vetoed to avoid the need to account for resolution effects, which are negligible for other, broader, resonances.
Existing knowledge of ${\cal B}(\Bp \to \jpsi h^+)$ and ${\cal B}(\jpsi \to h^{\prime +}h^{\prime -})$~\cite{PDG2020} is sufficient to correct for such differences in definition, which have an impact not larger than 1\%.

To determine the relative branching fraction of two modes, it is necessary to know the relative signal yields and efficiency of each.
By considering only ratios of these quantities, many sources of potentially large systematic uncertainty are rendered negligible.
However, the efficiency of each mode depends on its Dalitz-plot distribution, and for $\Bp \to \Kp\Kp\Km$ and $\Kp\pip\pim$ decays the most recent Dalitz-plot models~\cite{Garmash:2004wa,Lees:2012kxa,Garmash:2005rv,Aubert:2008bj} were obtained from analyses of significantly smaller samples than those in the current analysis.
To avoid a dominant systematic uncertainty due to lack of knowledge of the Dalitz-plot distributions, a model-independent approach is pursued whereby an efficiency correction is applied to each candidate depending on its Dalitz-plot position.

The remainder of the paper is organised as follows.
In Sec.~\ref{sec:Detector} the detector and simulation software is described.
The selection of signal candidates is discussed in Sec.~\ref{sec:Selection}, with the efficiency of these requirements, including the variations of the efficiency across the Dalitz plot of each of the final states, presented in Sec.~\ref{sec:Efficiency}.
In Sec.~\ref{sec:Fit-model} the simultaneous fit to the invariant-mass distributions of selected candidates is described, with emphasis on the various constraints that are imposed.
A detailed discussion of the evaluation of systematic uncertainties is presented in Sec.~\ref{sec:Systematics}, with the results and their correlations given in Sec.~\ref{sec:Results}.
A summary concludes the paper in Sec.~\ref{sec:Summary}.

\section{Detector and simulation}
\label{sec:Detector}

The \lhcb detector~\cite{LHCb-DP-2008-001,LHCb-DP-2014-002} is a single-arm forward spectrometer covering the \mbox{pseudorapidity} range $2<\eta <5$, designed for the study of particles containing \bquark or \cquark quarks.
The detector includes a high-precision tracking system consisting of a silicon-strip vertex detector surrounding the $pp$ interaction region~\cite{LHCb-DP-2014-001}, a large-area silicon-strip detector located upstream of a dipole magnet with a bending power of about $4{\mathrm{\,Tm}}$, and three stations of silicon-strip detectors and straw drift tubes~\cite{LHCb-DP-2013-003} placed downstream of the magnet.
The tracking system provides a measurement of the momentum, \ptot, of charged particles with relative uncertainty that varies from 0.5\% at low momentum to 1.0\% at 200\gevc.
The minimum distance of a track to a primary $pp$ collision vertex (PV), the impact parameter (IP), is measured with a resolution of $(15+29/\pt)\mum$, where \pt is the component of the momentum transverse to the beam, in\,\gevc.
Different types of charged hadrons are distinguished using information from two ring-imaging Cherenkov detectors~\cite{LHCb-DP-2012-003}.
Photons, electrons and hadrons are identified by a calorimeter system consisting of scintillating-pad and preshower detectors, an electromagnetic and a hadronic calorimeter.
Muons are identified by a system composed of alternating layers of iron and multiwire proportional chambers~\cite{LHCb-DP-2012-002}.

The online event selection is performed by a trigger~\cite{LHCb-DP-2012-004}, which consists of a hardware stage, based on information from the calorimeter and muon systems, followed by a software stage, in which all charged particles with $\pt>500\,(300)\mevc$ are reconstructed for 2011\,(2012) data.
At the hardware trigger stage, events are required to have a muon with high \pt or a hadron, photon or electron with high transverse energy deposited in the calorimeters.
For hadrons, the transverse energy threshold is $3.5\gev$.
The software trigger requires a two-, three- or four-track vertex with significant displacement from any PV.
At least one charged particle must have transverse momentum $\pt > 1.6\gevc$ and be inconsistent with originating from a PV.
A multivariate algorithm~\cite{BBDT} is used for the identification of displaced vertices consistent with the decay of a \bquark hadron.

In the offline selection, trigger signals are associated with reconstructed particles.
Selection requirements can therefore be made on the trigger output and on whether the decision was due to the signal candidate, other particles produced in the $pp$ collision, or a combination of both.
In this analysis it is required that the hardware trigger decision is due to either clusters in the hadronic calorimeter created by one or more of the final-state particles, or only by particles produced in the $pp$ bunch crossing not involved in forming the \B\ candidate.

Simulation is used to model the effects of the detector acceptance and the selection requirements.
In the simulation, $pp$ collisions are generated using \pythia~\cite{Sjostrand:2007gs,*Sjostrand:2006za} with a specific \lhcb configuration~\cite{LHCb-PROC-2010-056}.
Decays of unstable particles are described by \evtgen~\cite{Lange:2001uf}, in which final-state radiation is generated using \photos~\cite{Golonka:2005pn}.
The interaction of the generated particles with the detector, and its response,  are implemented using the \geant  toolkit~\cite{Allison:2006ve, *Agostinelli:2002hh} as described in Ref.~\cite{LHCb-PROC-2011-006}.

\section{Selection of signal candidates}
\label{sec:Selection}

The procedure to select signal candidates is similar to those used in previous LHCb analyses of $\Bp \to h^+h^{\prime +}h^{\prime -}$ decays~\cite{LHCb-PAPER-2013-027,LHCb-PAPER-2013-051,LHCb-PAPER-2014-044,LHCb-PAPER-2018-051,LHCb-PAPER-2019-017,LHCb-PAPER-2019-018}, but is optimised for the set of relative branching fraction measurements of this analysis.
A loose set of initial requirements is applied, and particle identification (PID) requirements are imposed to reject background with misidentified final-state particles.
A multivariate algorithm (MVA) is used to distinguish signal from combinatorial background.
Further specific requirements are applied to remove potentially large background sources from candidates where two of the final-state particles originate from a charm- or beauty-meson decay.

The initial selection includes requirements on the quality of each of the three tracks comprising the signal candidate.
They are required to be displaced from all PVs, as quantified through the variable \chisqip, which is the difference in the vertex-fit \chisq of a given PV reconstructed with and without the particle under consideration.
The three tracks must form a common, good-quality vertex, and have invariant mass within a broad window of the known \Bp\ mass~\cite{PDG2020}.
The \B~candidate is associated to the PV with which it forms the minimum \chisqip\ value, which must be below a certain threshold, and the \B-candidate momentum must be aligned with the vector between its production and decay vertices.
The \B decay vertex must be displaced significantly from its associated PV.
Requirements are also imposed on the \ptot\ and \pt\ of the \B~candidate and of the individual tracks.
Variables used subsequently in the analysis are obtained from a kinematic fit to the decay~\cite{Hulsbergen:2005pu} in which the tracks are constrained to a common vertex. 
For the computation of Dalitz-plot variables, the \B~candidate is additionally constrained to have the known \Bp mass~\cite{PDG2020}.

Information from the ring-imaging Cherenkov detectors is combined with tracking information to obtain variables that quantify how likely a given track is to be caused by either a pion or a kaon~\cite{LHCb-DP-2012-003}.
Disjoint regions in the plane formed by these two variables are used to separate tracks that are likely to originate from kaons and unlikely to come from pions and vice versa.
For each of the four final states, requirements on these PID variables are imposed to reduce the potential cross-feed background from misidentification of the other modes.
Optimal requirements are evaluated by considering the figure of merit $N_{\rm S}/\sqrt{N_{\rm S}+N_{\rm B\,cf}}$, where $N_{\rm S}$ and $N_{\rm B\,cf}$ are the expected signal and cross-feed background yields for each case.
The relative sizes of $N_{\rm S}$ and $N_{\rm B\,cf}$ depend on the branching fractions of the four signal modes, which are taken from previous measurements~\cite{PDG2020}, as well as efficiencies and misidentification rates.
These are determined from data control samples of \decay{\Dstarp}{\Dz(\to\Km\pip)\pip} decays~\cite{LHCb-DP-2012-003}, weighted to reproduce the \ptot\ and $\eta$ distributions of signal tracks, and --- since the PID performance depends on the detector occupancy --- the number of reconstructed tracks in the $pp$ bunch crossing.
Requirements on the ranges of these three variables are applied to ensure reliable performance of the PID calibration procedure.
Tracks are also required to not have any associated signal in the muon detectors.
For the $\Bp \to \pip\Kp\Km$ channel, the expected significant cross-feed background from partially reconstructed $B \to \Kp\pip\pim X$ and $\Kp\Kp\Km X$ decays, where $X$ denotes any additional particles, is accounted for by doubling the value of $N_{\rm B\,cf}$ from that obtained considering the three-body $\Bp$ decays only.
A baseline set of PID requirements is applied, in the cases where the optimisation procedure returns loose values, to ensure that no candidate can be selected in more than one of the final states under consideration.
The outcome of this procedure is a set of requirements that, after further tightening in certain regions of phase space as described below,
corresponds to the efficiencies and misidentification rates given in Table~\ref{table:pid}.

\begin{table}[!tb]
  \centering
  \caption{
    Probability (\%), due to the particle identification requirements, for each of the four signal modes to be correctly identified, or to form a cross-feed background to one of the other final states.
    Empty entries correspond to values below $0.05\,\%$.
    The decays $\Bp \to \pip\Kp\Km$ and $\Bp \to \Kp\pip\pim$ can, through both $\Kp \to \pip$ and $\pip \to \Kp$ misidentification, appear as a cross-feed background in the correct final state with probabilities of below $0.05\,\%$ and $0.4\,\%$, respectively.
  }
  \label{table:pid}
  \begin{tabular}{l|cccc}
    \hline
    \hline
    Decay & \multicolumn{4}{c}{Reconstructed final state} \\
    & $\Kp\Kp\Km$ & $\pip\Kp\Km$ & $\Kp\pip\pim$ & $\pip\pip\pim$ \\
    \hline
    $\Bp \to \Kp\Kp\Km$ & 77.1 & 0.7 & 0.3 & -- \\
    $\Bp \to \pip\Kp\Km$ & 6.5 & 42.1 & 4.5 & -- \\
    $\Bp \to \Kp\pip\pim$ & 0.5 & 1.0 & 65.9 & 5.8 \\
    $\Bp \to \pip\pip\pim$ & -- & -- & 3.4 & 70.2 \\
    \hline
    \hline
  \end{tabular}
\end{table}

Variables that provide good discrimination between signal and combinatorial background without introducing significant distortions into the \B-candidate mass or Dalitz-plot distributions, are identified for inclusion in the MVA.
In order of discriminating power, these are:
the pointing angle, which characterises how well the \B-candidate momentum aligns with the vector from the associated PV to the \B decay vertex;
the $\pt$ asymmetry, which quantifies the isolation of the \B candidate through the \pt\ asymmetry between itself and other tracks within a cone around its flight direction~\cite{LHCb-PAPER-2012-001};
the distance between the \B-candidate production and decay vertices, divided by its uncertainty;
the \chisq\ of the \B-candidate vertex;
the \chisqip\ of the track with the largest \pt\ out of the three that form the \B~candidate;
the \ptot\ of the same track;
the \chisqip\ of the \B candidate.
These variables are distributed almost identically for all signal modes, justifying the use of a single MVA.
The distributions of all input variables, and the MVA output, are confirmed to agree well between data and simulation, where the data distributions are obtained from the $\Bp \to \Kp\Kp\Km$ sample with background subtracted using weights obtained from a fit to the \B-candidate mass distribution~\cite{Pivk:2004ty}.

The combination of variables into the MVA is implemented with the \neurobayes\ package~\cite{Feindt:2006pm}.
The MVA is trained to discriminate between a signal sample, taken from simulation, and a background sample obtained from the data sideband with \B-candidate mass values significantly above the \Bp mass.
Since  the decay $\Bp \to \pip\Kp\Km$ is the most challenging of the four modes to separate from background, the training is performed with both signal and background samples corresponding to that mode, with initial selection and PID requirements applied.
A requirement on the output of the MVA is optimised by considering the figure of merit $N_{\rm S}/\sqrt{N_{\rm S}+N_{\rm B\,cb}}$, where $N_{\rm B\,cb}$ is the expected combinatorial background yield in the signal region $\left[5240, 5320\right]\mevcc$.

Background from $\Bp \to \Dzb h^+$ decays, with $\Dzb \to \Kp\pim$, $\Kp\Km$ or $\pip\pim$, passes the selection requirements for the correctly reconstructed final state in large numbers, since the \Dzb\ lifetime is sufficiently small that the three tracks can still form a good \B-candidate vertex.
This background is vetoed by removing any candidate with one of the corresponding two-body invariant masses in the region $\left[1830, 1890\right] \mevcc$.
Such decays can still cause background when final-state particles are misidentified.
Tighter PID requirements are imposed when one of the two-body invariant masses of oppositely charged final-state particles is in the range $\left[1890, 2000\right] \mevcc$  for $\pi \to K$ misidentification or $\left[1700, 1850\right] \mevcc$ for $K \to \pi$ misidentification.
These requirements reduce most misidentified charm background components to negligible levels with minimal impact on the signal efficiency.

The so-called partially combinatorial background, where a two-body \B-meson decay is combined with a random track, can populate the \B-candidate invariant-mass region at values above the signal peaks.
The shape of such background can be hard to model in the \B-candidate invariant-mass fit, introducing a potential source of systematic uncertainty on the signal yield.
Therefore, candidates that may contain $\Bz \to \Kp\pim$, $\Bz \to \pip\pim$ or $\Bs \to \Kp\Km$ decays are removed by vetoing the two-body invariant-mass ranges $\left[5220, 5320\right] \mevcc$ and $\left[5300, 5400\right] \mevcc$ under the appropriate hypotheses.
Partially combinatorial background with misidentification of final-state particles has a \B-candidate mass distribution that is sufficiently broad that it can be absorbed into the combinatorial background component.
Similarly, the impact of partially combinatorial background from the suppressed $\Bd \to \Kp\Km$, $\Bs \to \Km\pip$ and $\Bs \to \pip\pim$ decays~\cite{LHCb-PAPER-2016-036} is negligible.

After all selection requirements are imposed, a small fraction of selected $pp$ bunch crossings, ranging from $0.2\,\%$ for the $\Kp\Kp\Km$ final state to $2.4\,\%$ for $\pip\pip\pim$, contain more than one \B candidate.
In such cases, only the candidate with the highest MVA output value is retained.
The systematic uncertainty associated with this procedure is negligible.

\section{Signal efficiency}
\label{sec:Efficiency}

The total signal efficiency, $\epsilon^{\rm tot}$, can be expressed in terms of factorising components,
\begin{equation}
  \label{eq:effcomponents}
  \epsilon^{\rm tot} = \epsilon^{\rm sel+geom} \times \epsilon^{\rm PID}\,,
\end{equation}
where $\epsilon^{\rm sel+geom}$ includes the effects of the geometrical efficiency of the \lhcb\ detector and of both online and offline selection requirements, and $\epsilon^{\rm PID}$ is the PID efficiency for candidates that have passed the selection requirements.
The former can be evaluated quite reliably from simulation, although small data-driven corrections are applied, while the latter is obtained from control samples.
As explained in Sec.~\ref{sec:Introduction}, the variation of the efficiency across the phase space, or Dalitz plot, of each decay, must be accounted for.
It is convenient to do so using the so-called square Dalitz plot (SDP) representation of the phase space, since this provides greater granularity in regions close to the edges of the regular Dalitz plot where resonances tend to populate and where the efficiency variation tends to be larger.  
Moreover, the SDP definition in terms of two variables $\mpr$ and $\thetapr$, each of which is bounded in the range $\left[ 0, 1\right]$, aligns a rectangular grid with the edges of the phase space, avoiding edge effects associated with rectangular binning of the regular Dalitz plot.
The variable $\mpr$ is a transformation of the invariant mass of two of the three final-state particles, while $\thetapr$ is a transformation of the helicity angle associated with that pair, \ie\ the angle between the momentum of one of the pair and the third particle in the rest frame of the pair.
The explicit definitions are~\cite{Back:2017zqt}
\begin{eqnarray}
\label{eq:mPrimeFull}
\mpr & = & \frac{1}{\pi}\arccos{\left(2 \frac{m_{ij}-(m_i+m_j)}{m_{\B}-(m_i+m_j+m_k)} -1\right)}\,, \\
\label{eq:thPrimeFull}
\thetapr & = & \frac{1}{\pi} \left( \frac{m^2_{ij}(m^2_{jk}-m^2_{ij}) - (m^2_j-m^2_i)(m^2_{\B}-m^2_k)}{\sqrt{(m^2_{ij}+m^2_i-m^2_j)^2-4m^2_{ij}m^2_i} \sqrt{(m^2_{\B}-m^2_k-m^2_i)^2-4m^2_{ij}m^2_k}} \right)\,,
\end{eqnarray}
where the ordering of the particles used in the analysis is given in Table~\ref{tab:sqDPdef}, $m_{\alpha}$ is the mass of the particle labelled $\alpha$ and $m_{\alpha\beta}$ is the two-body invariant mass of particles $\alpha$ and $\beta$.
For decays with two identical particles, \ie \BuToKKK and \BuTopipipi, the SDP is folded along the line $\thetapr = 0.5$, making the initial ordering, \ie\ which of the two identical particles is $i$ and which is $j$, irrelevant.
The simulated samples of signal decays used in the analysis to determine $\epsilon^{\rm sel+geom}$ are generated with uniform density in these SDP coordinates.

\begin{table}[!tb]
  \centering
  \caption{
    Ordering of final-state particles used in definitions of the SDP variables.
  }
  \label{tab:sqDPdef}
  \begin{tabular}{lccc}
    \hline\hline
    Decay & $i$ & $j$ & $k$ \\
    \hline
    $\Bp \to \Kp\Kp\Km$ & \Kp & \Kp & \Km \\
    $\Bp \to \pip\Kp\Km$ & \pip & \Km & \Kp \\
    $\Bp \to \Kp\pip\pim$ & \pip & \pim & \Kp \\
    $\Bp \to \pip\pip\pim$ & \pip & \pip & \pim \\
    \hline\hline
  \end{tabular}
\end{table}

The impact of the hardware trigger is a potentially significant source of discrepancy between data and simulation in the evaluation of $\epsilon^{\rm sel+geom}$.
Corrections to the simulation are applied for two mutually exclusive subsamples of the selected candidates.
The first includes candidates that are triggered at hardware level by clusters in the hadronic calorimeter created by one or more of the final-state signal particles, and the second contains those triggered only by other particles produced in the $pp$ bunch crossing.
For the first subsample, a correction is calculated from the probability of an energy deposit in the hadronic calorimeter to fire the trigger, evaluated from calibration data samples as a function of particle type (kaon or pion), charge, dipole magnet polarity, transverse energy and position in the calorimeter.
In the second subsample, the simulation is weighted so that the rates of the different categories of hardware trigger (hadron, muon, dimuon, electron, photon) match those observed in data.
As described in Sec.~\ref{sec:Systematics}, the former of these corrections has a non-negligible impact on the results, while the effect of the latter is smaller.
Additional small corrections are applied to the simulation to ensure that the tracking efficiency~\cite{LHCb-DP-2013-002}, and the kinematic ($\pt, \eta$) distributions of selected \B\ mesons match those of data.

The PID efficiency is calculated, in the same way as described above for the optimisation of the PID requirements, from calibration samples.
The efficiencies for each final-state particle are parameterised in terms of their total and transverse momentum, and the number of tracks in the event, and these are multiplied to form the overall efficiency $\epsilon^{\rm PID}$.

\begin{figure}[!tb]
  \centering
  \includegraphics[width=0.45\textwidth]{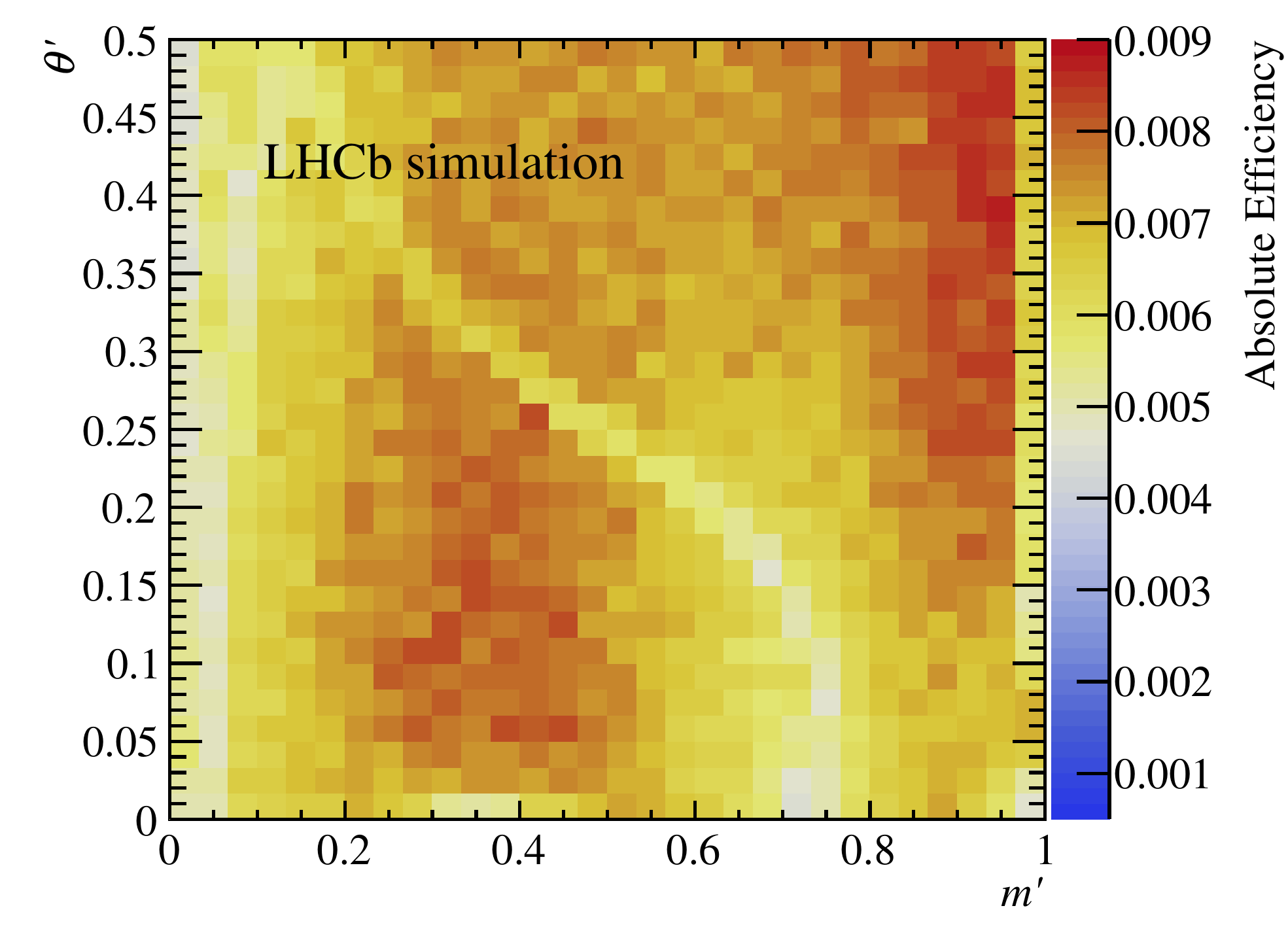}
  \includegraphics[width=0.45\textwidth]{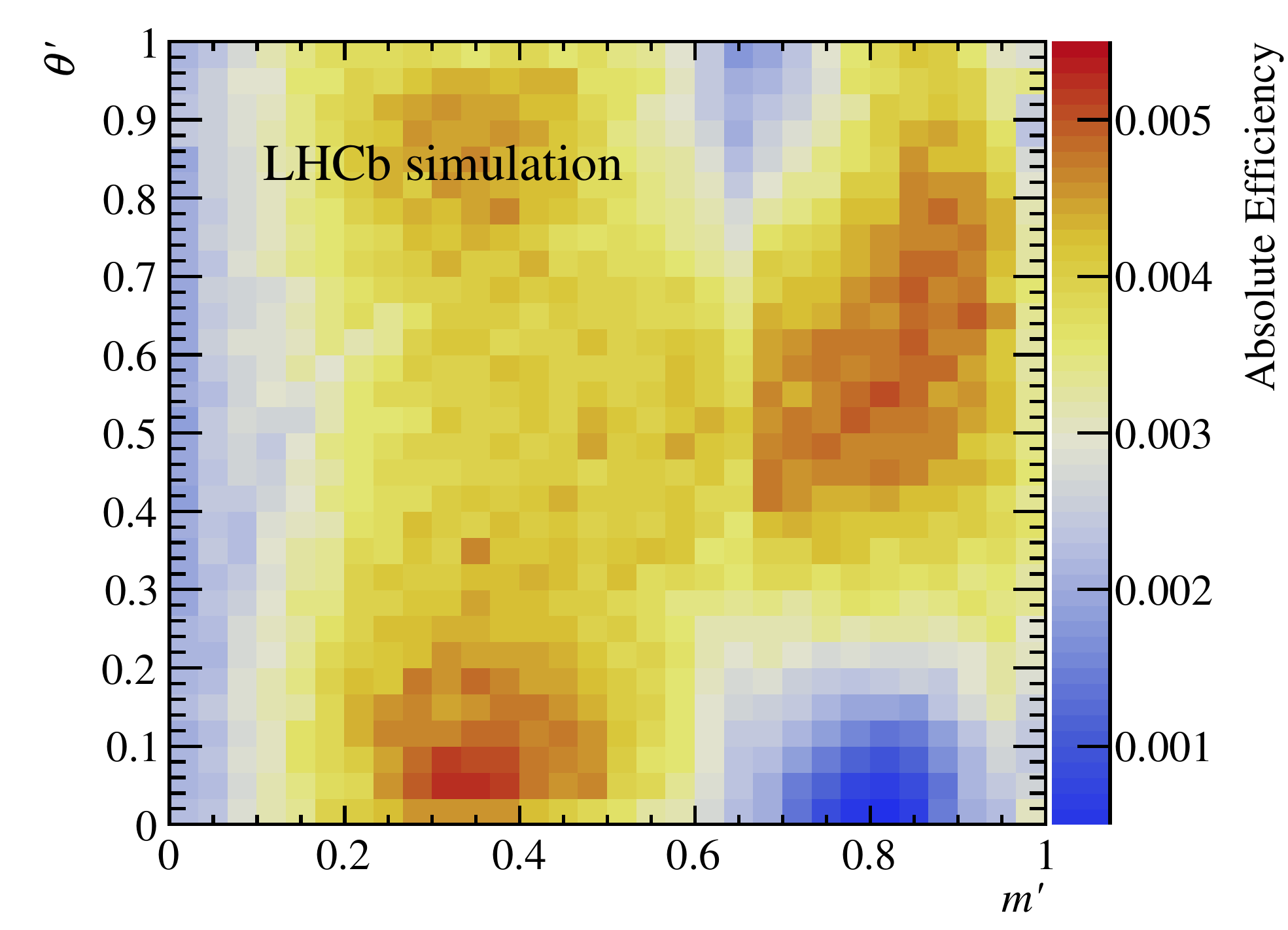}

  \includegraphics[width=0.45\textwidth]{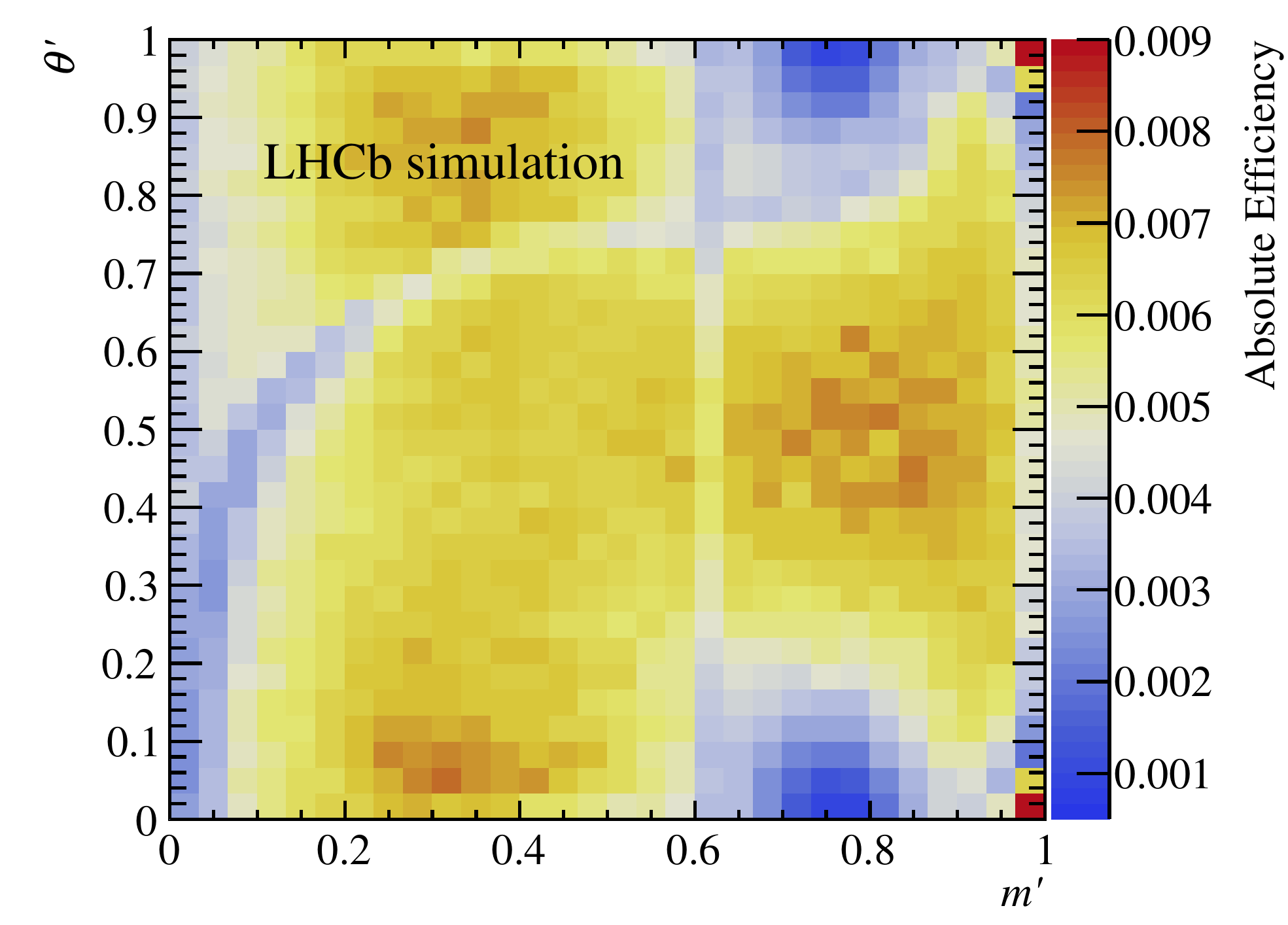}
  \includegraphics[width=0.45\textwidth]{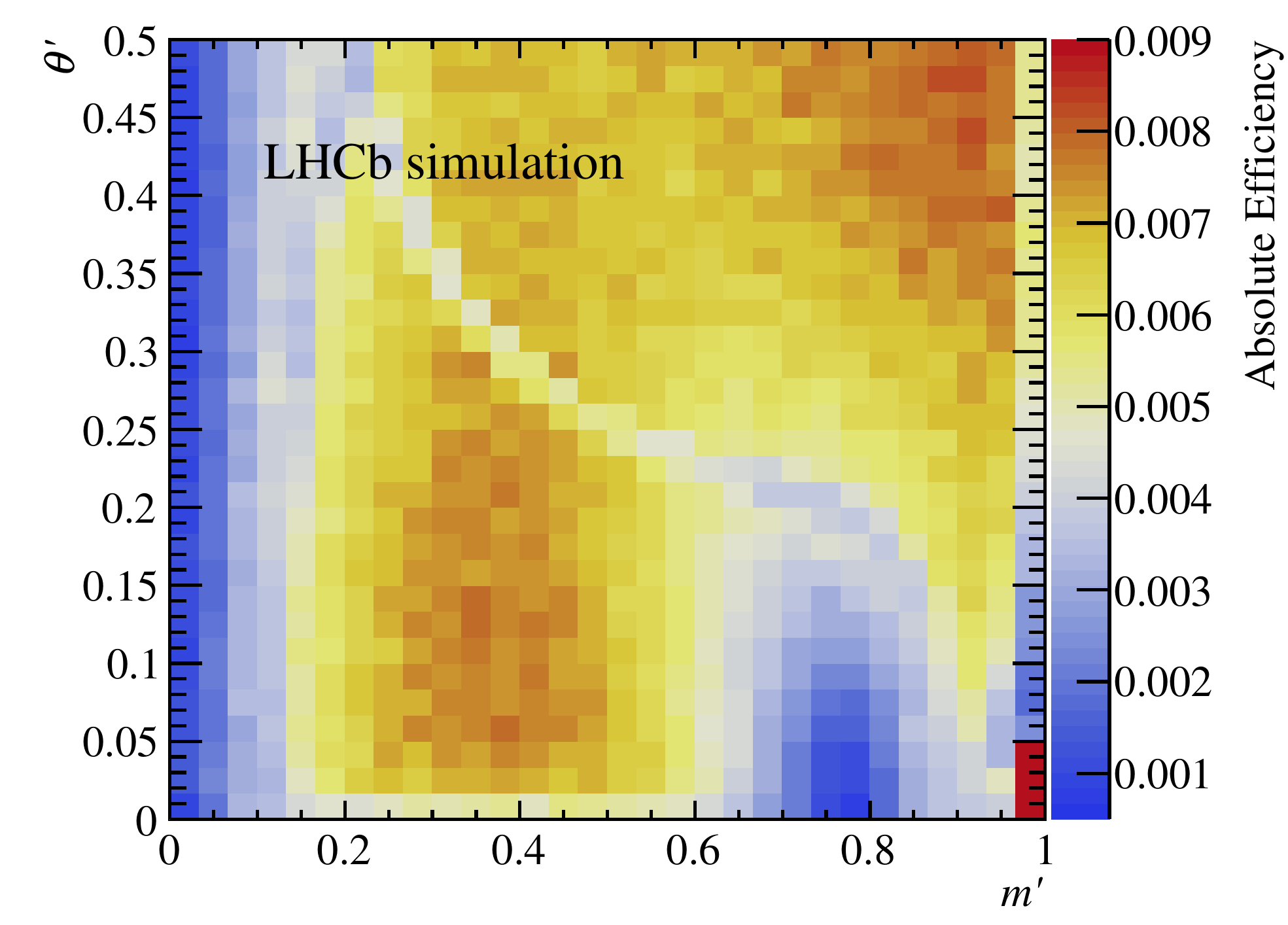}
  \caption{Total efficiency, $\epsilon^{\rm tot}$, as a function of SDP position for (top~left)~\BuToKKK, (top~right)~\BuToKpiK, (bottom~left)~\BuToKpipi, and (bottom~right)~\BuTopipipi.}
  \label{fig:totaleffs}
\end{figure}

The total efficiency, $\epsilon^{\rm tot}$, is shown in Fig.~\ref{fig:totaleffs} as a function of SDP position for the four signal modes, with all selection requirements except the charm vetoes applied.
Bands in the phase space are nevertheless visible around the charm-meson mass due to the tighter PID requirements applied in these regions.
For example, the depleted region in $\epsilon^{\rm tot}$ for \BuToKKK\ decays is due to tightened PID requirements to remove $\Bp\to\Dzb(\to\Kp\pim)\Kp$ decays with $\pim \to \Km$ misidentification.
The choice of $30\times30$ bins in these efficiency maps is made so that the minimum bin content remains above 10 and hence the efficiency in each bin is determined with reasonably small uncertainty, although some fluctuations are visible at the edges, and particularly the corners, of the SDP.
These fluctuations occur where the Jacobian of the transformation from conventional to SDP coordinates takes extreme values, and hence affect modes with final-state pions more than kaons.

Since candidate-by-candidate efficiency corrections are applied in the evaluation of the relative branching fractions, the impact of charm vetoes that completely remove regions of phase space is accounted for separately.
The veto efficiencies are determined by generating ensembles of samples according to the most recent Dalitz-plot models of the signal modes~\cite{Aubert:2008bj,Lees:2012kxa,LHCb-PAPER-2018-051,LHCb-PAPER-2019-017,LHCb-PAPER-2019-018}, and evaluating the impact of the veto.
Each sample contains a number of decays sampled from a Poisson distribution with mean corresponding to the signal yield in the analysis where the model was determined, and the corresponding uncertainties are estimated from the spread of veto efficiency values in the ensemble.
The efficiencies obtained for each channel, $\epsilon^{\rm veto}$, are given in Table~\ref{tab:vetoEfficiencies}.

\begin{table}[!tb]
\caption{Charm veto efficiencies as determined from samples of the four signal modes generated according to the most recent Dalitz-plot models.}
\label{tab:vetoEfficiencies}
\centering
\begin{tabular}{lr@{$\,\pm\,$}l}
\hline
\hline
Decay & \multicolumn{2}{c}{$\epsilon^{\rm veto}$ (\%)} \\
\hline
\hline
\BuToKKK        & 97.52 & 0.22 \\
\BuToKpiK       & 98.41 & 0.21 \\
\BuToKpipi      & 97.92 & 0.19 \\
\BuTopipipi     & 98.05 & 0.10 \\
\hline
\hline
\end{tabular}
\end{table}

\section{\boldmath \B-candidate invariant-mass fit}
\label{sec:Fit-model}

A simultaneous unbinned extended maximum-likelihood fit is performed to the four \B-candidate invariant-mass distributions, in the range $\left[5100, 5500\right]\mevcc$, to determine the yields of the signal components.
The fit model includes components for signal, cross-feed from misidentified three-body \B decays, partially reconstructed background and combinatorial background.

The signal mass distributions are modelled as the sum of two Crystal Ball functions~\cite{Skwarnicki:1986xj}, with a common peak position and width, and tails to opposite sides of the peak.
The shape parameters of the double Crystal Ball function are determined from fits to simulation and then fixed in the data fit, with the exception of an offset to the peak position and a scaling factor of the width.
These two parameters, shared by all four modes, are both left free to vary in the fit to data to account for small differences between data and simulation.

All possible cross-feed background contributions from one $\Bp \to h^+h^{\prime +}h^{\prime -}$ decay to another, or to itself, with single or double misidentification are accounted for in the fit.
The shapes are described empirically with the sum of two Crystal Ball functions, with parameters obtained from simulated samples weighted to reproduce the underlying Dalitz-plot distributions~\cite{Aubert:2008bj,Lees:2012kxa,LHCb-PAPER-2018-051,LHCb-PAPER-2019-017,LHCb-PAPER-2019-018} and with per-track data-calibrated PID efficiencies applied.
The peak positions and widths of these shapes are adjusted, in the fit to data,  by the same offset and scale factor as the signal functions.
Other potential sources of similar background, involving misidentified three-body \bquark-hadron decays such as $\Xibbarp \to h^+h^{\prime +}\antiproton$~\cite{LHCb-PAPER-2016-050} are found to have negligible contribution.

The sources of partially reconstructed background differ between the four final states considered.
All include a component from four-body charmless $\Bp$ and $\Bz$ decays with an additional soft neutral or charged pion that is not reconstructed.
The shapes of these, and all partially reconstructed background components, are modelled with ARGUS functions~\cite{Albrecht:1990am}, where the threshold is fixed to the known difference between the \B-meson and pion masses~\cite{PDG2020}, convolved with a Gaussian resolution function with width of the corresponding signal mode.
The shape parameters are fixed to the values obtained from fitting simulated samples of the background.

For all modes except $\Bp\to\Kp\Kp\Km$, there is significant background from \mbox{$\Bs\to\Dsm\pip$} decays, with subsequent $\Dsm$ decay to the corresponding pair of particles plus an additional soft pion that is not reconstructed.
The shapes of these components differ from those of the corresponding charmless four-body decays because of differences in the momentum distributions of the missing pion.
The same parametric functions are used as for the charmless four-body decays, but with parameters determined independently from appropriate simulation samples.

The $\pip\Kp\Km$ final state has a further source of partially reconstructed background through $\Bs\to \pip\Kp\Km\pim$ decays.
The latest study of this process~\cite{LHCb-PAPER-2017-048} reveals that it is composed of a mixture of $K\pi$ resonances, rather than being dominated by the $\Bs \to \Kstar(892)^0\Kstarb(892)^0$ decay, so a data-driven approach is used to determine the shape of this component.

The $\Kp\pip\pim$ final state contains background from $\Bp \to \etapr\Kp$ with $\etapr \to \pip\pim\gamma$ decays.
The ARGUS function shape parameter is determined from a fit to a sample of simulation weighted to reproduce the appropriate $\pip\pim$ invariant-mass shape~\cite{Ablikim:2017fll}.
The threshold parameter is fixed to the peak value of the $\Bp\to\Kp\pip\pim$ signal decay including, in the fit to data, the offset.

Background to the $\Bp\to\pip\pip\pim$ decay from misidentified $\Bp \to \Dzb(\to \Kp\pim)\pip$ decays remains at non-negligible level after the PID requirements.
This is modelled in the fit with an ARGUS function convolved with a Gaussian resolution, with parameters determined from a fit to simulation, in a similar way as for the partially reconstructed background.
Misidentified $\Bp \to \Dzb(\to \Kp\pim)\pip$ decays are also a source of background in the $\pip\Kp\Km$ final state, but this is found to be readily absorbed by other fit components and is therefore not included explicitly.
The combinatorial background in each final state is described by an exponential function.

The free parameters of the fit are the four signal yields, the common offset and scale factor of the signal shape functions, the four combinatorial background yields and their associated exponential shape parameters, one partially reconstructed background yield for each of the $\Kp\Kp\Km$, $\pip\Kp\Km$ and $\pip\pip\pim$ final states and two for the $\Kp\pip\pim$ channel.
All misidentified background yields are constrained, within uncertainty, to their expected levels based on the signal yields in the corresponding correctly identified final states and the known misidentification probabilities, as given in the off-diagonal elements of Table~\ref{table:pid}.
For background from misidentified $\Bp \to \Dzb(\to \Kp\pim)\pip$ decays, the known branching fraction, relative to those of the signal channels, also enters the calculation of the constraint.
Similarly, the relative yields of the different sources of partially reconstructed background in the $\pip\Kp\Km$ and $\pip\pip\pim$ final states, and of the $\Bp\to\etapr\Kp$ background to the $\Kp\pip\pim$ final state, are constrained to their expected values.

\begin{figure}[!tb]
  \centering
  \includegraphics[width=0.45\textwidth]{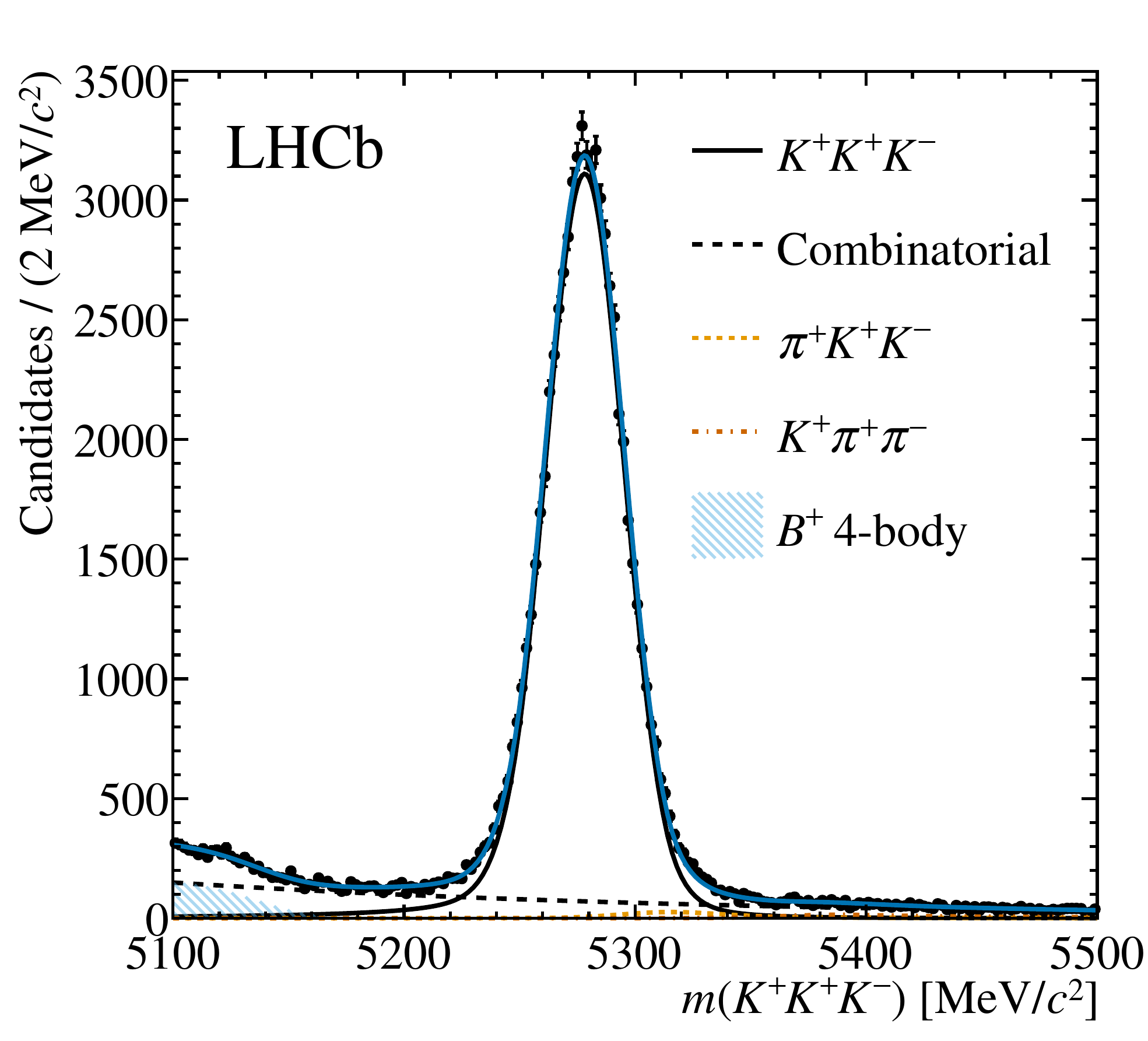}
  \includegraphics[width=0.45\textwidth]{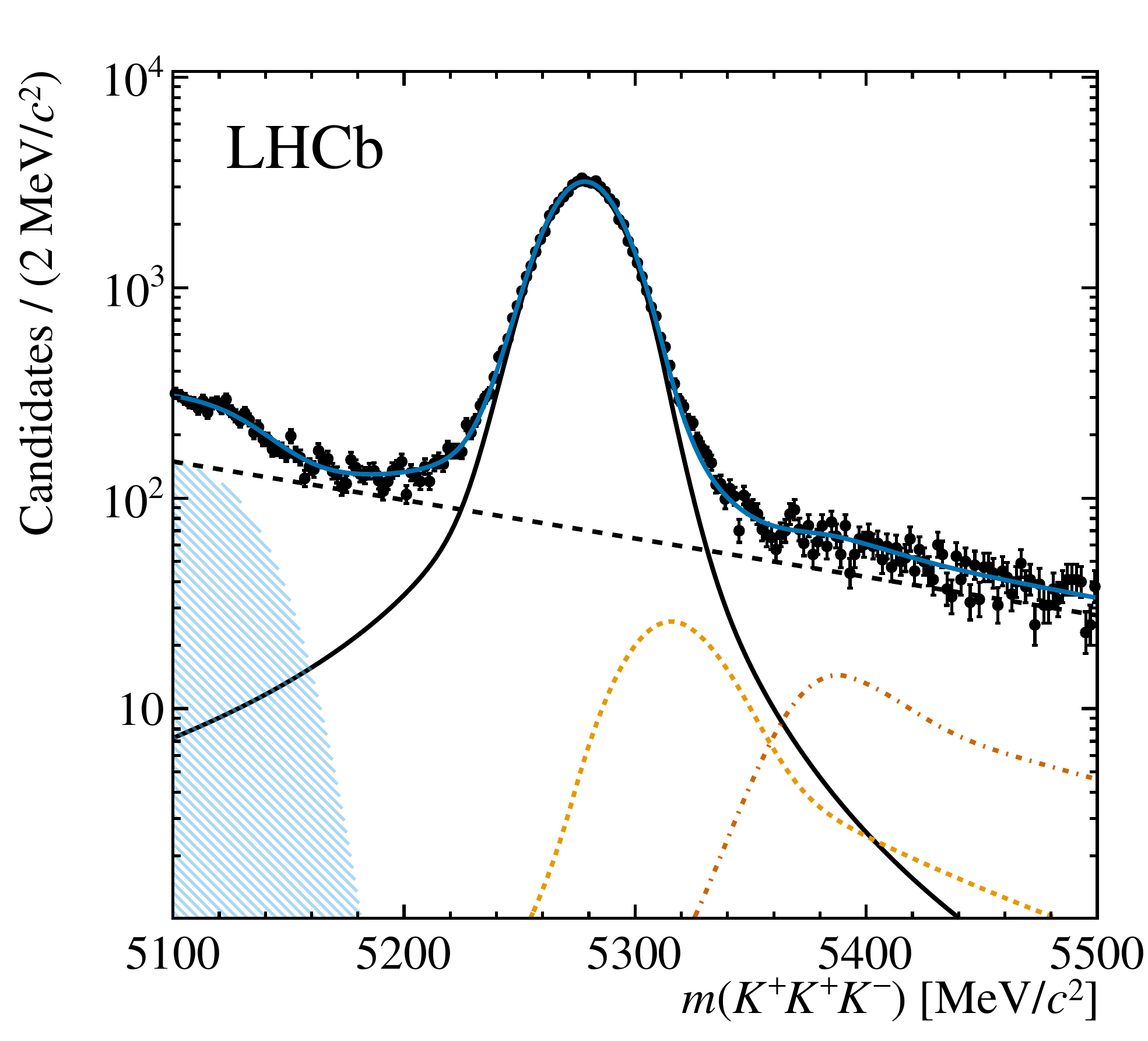}
  \includegraphics[width=0.45\textwidth]{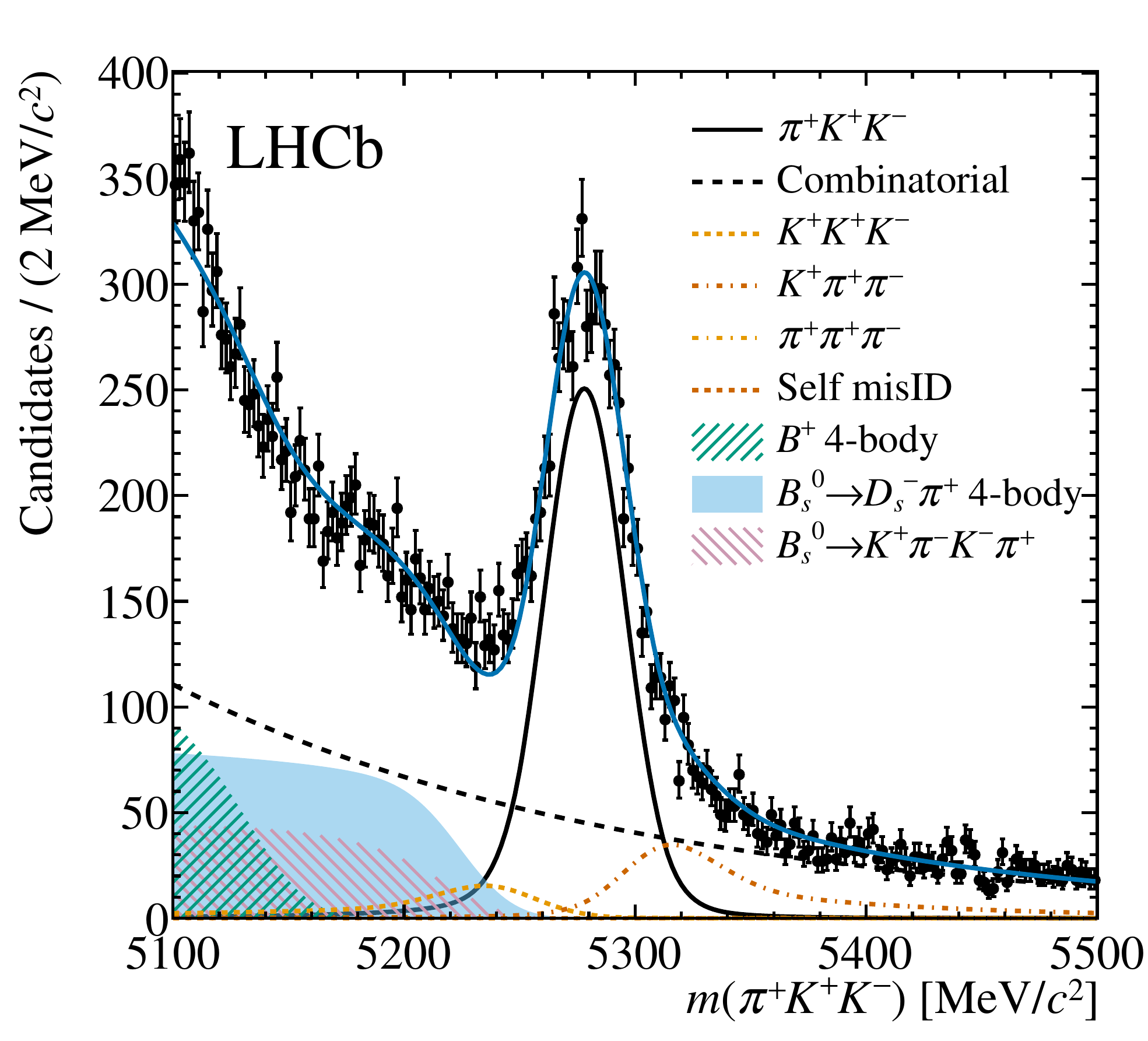}
  \includegraphics[width=0.45\textwidth]{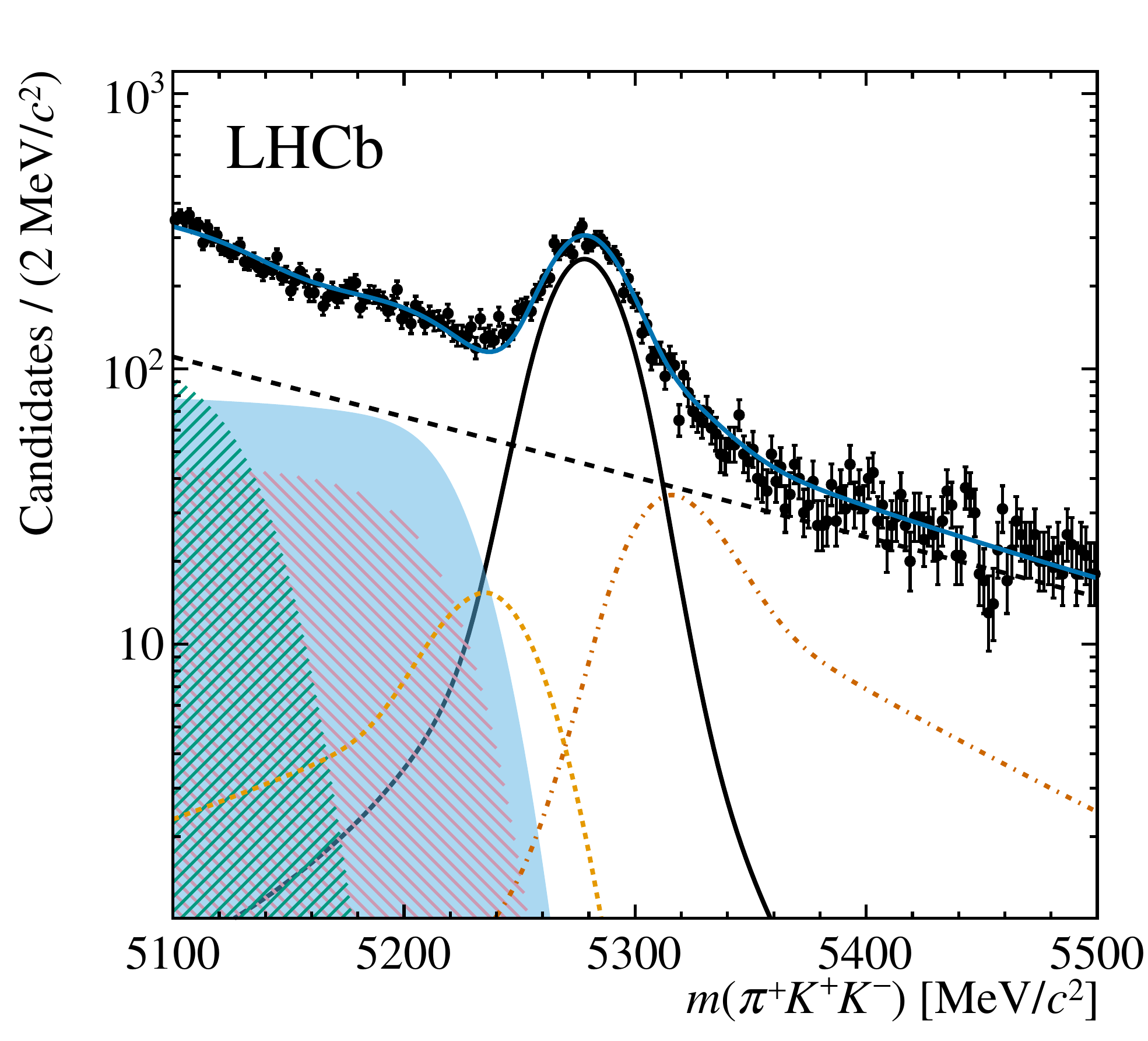}
  \caption{Invariant-mass distributions of the (top)~\BuToKKK\ and (bottom)~\BuToKpiK\ candidates compared to the results of the simultaneous fit with (left)~linear and (right)~logarithmic $y$-axis scales.}
  \label{fig:simfit_results1}
\end{figure}

\begin{figure}[!tb]
  \centering
  \includegraphics[width=0.45\textwidth]{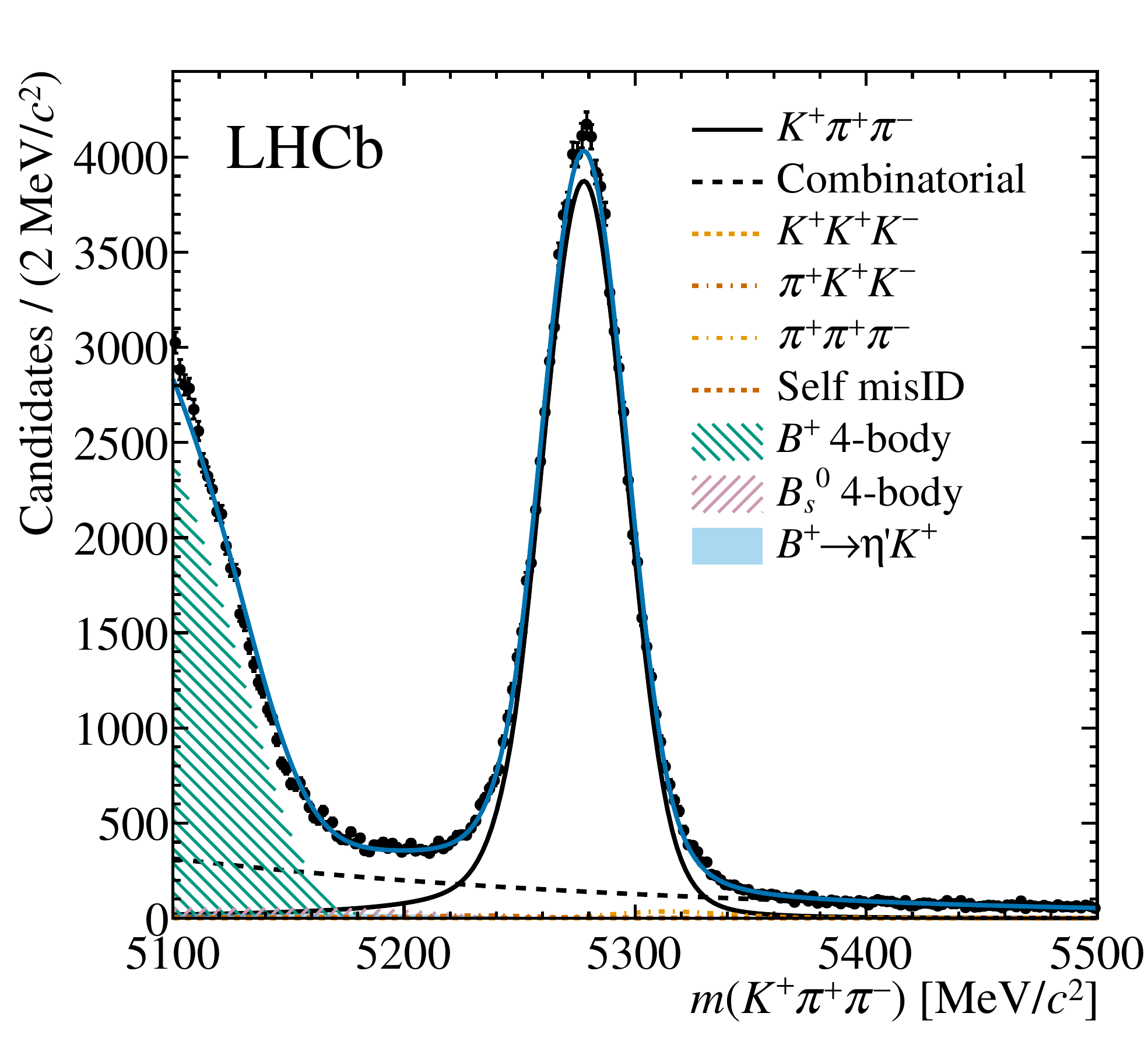}
  \includegraphics[width=0.45\textwidth]{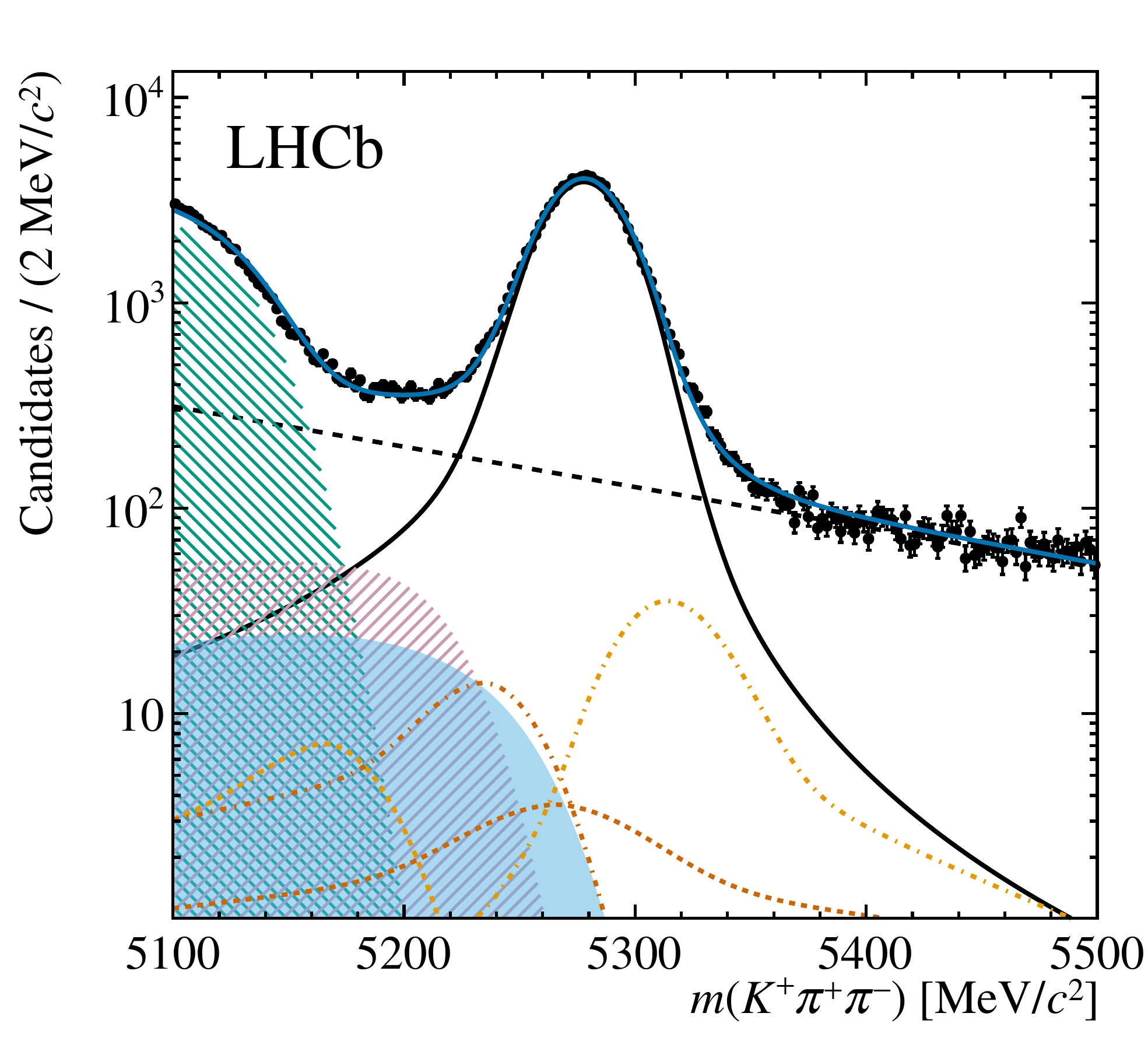}
  \includegraphics[width=0.45\textwidth]{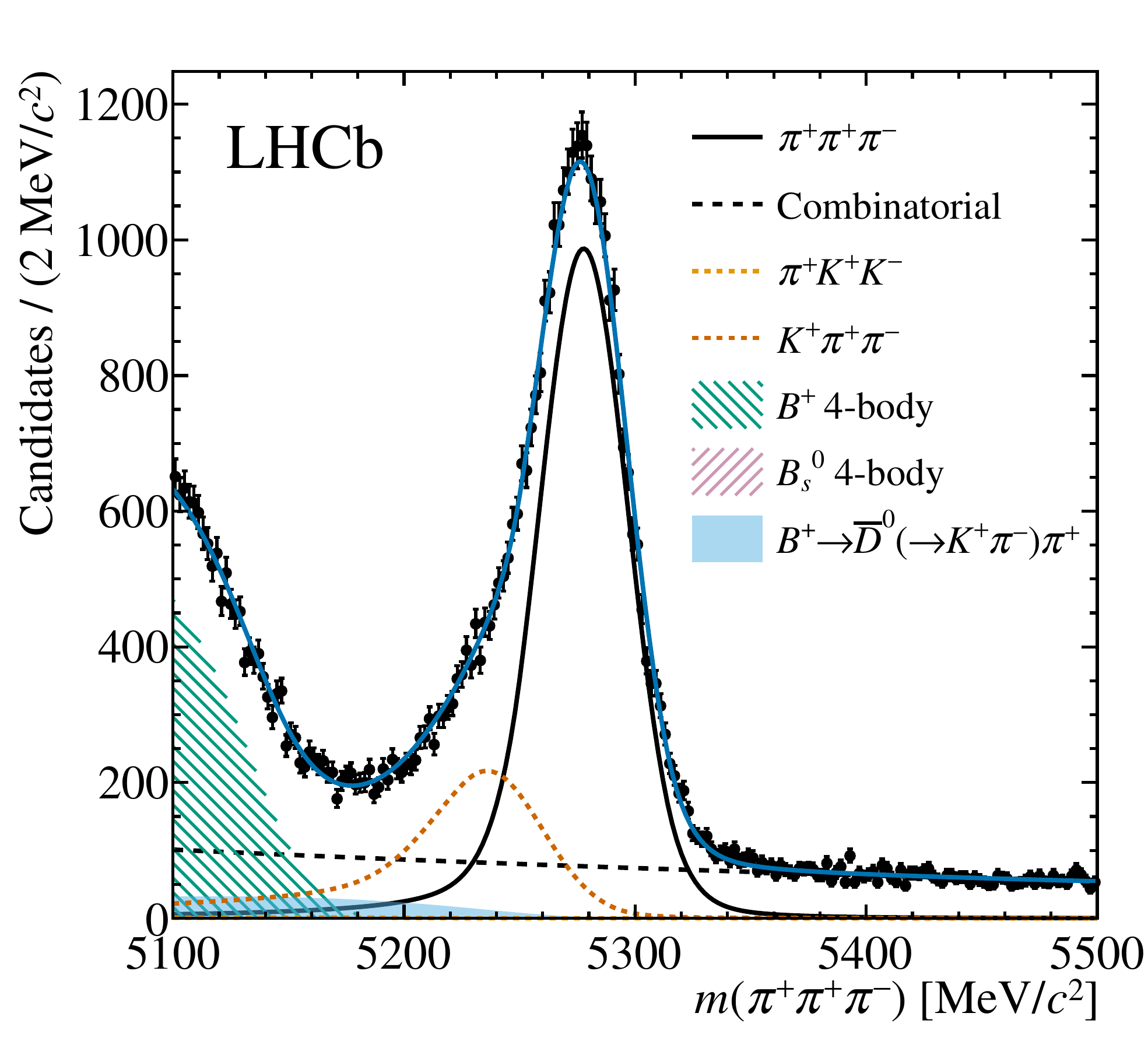}
  \includegraphics[width=0.45\textwidth]{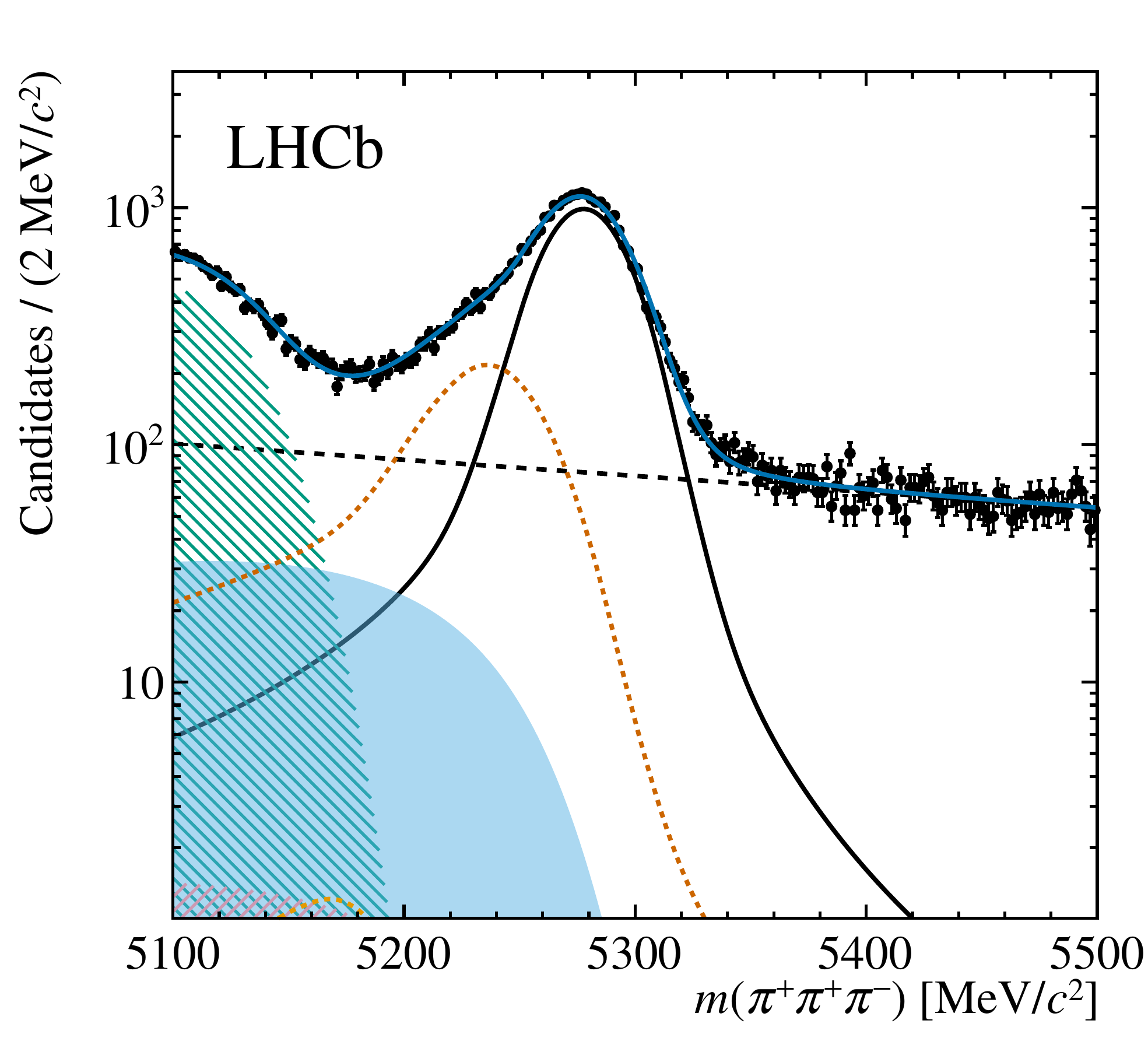}
  \caption{Invariant-mass distributions of the
  (top)~\BuToKpipi\ and (bottom)~\BuTopipipi\ candidates compared to the results of the simultaneous fit with (left)~linear and (right)~logarithmic $y$-axis scales.}
  \label{fig:simfit_results2}
\end{figure}

The invariant-mass distributions $m( h^+h^{\prime +}h^{\prime -})$ for selected candidates in all four signal modes together with the fit projections  are shown in Fig.~\ref{fig:simfit_results1} for the $\Kp\Kp\Km$ and $\pip\Kp\Km$ final states and in Fig.~\ref{fig:simfit_results2} for the $\Kp\pip\pim$ and $\pip\pip\pim$ final states.
The signal yields are given in Table~\ref{table:signal_yields}.
There is good agreement of the fit model with the data in all four final states, with some potential small residual discrepancies accounted for as sources of systematic uncertainty.
The stability of the fit is investigated with pseudoexperiments, and the signal yields are found to be unbiased within the statistical precision of the ensemble.

\begin{table}[!tb]
  \caption{Fitted signal yields and associated statistical uncertainties.}
\label{table:signal_yields}
\centering
\begin{tabular}{lr@{$\,\pm\,$}l}
\hline
\hline
Decay & \multicolumn{2}{c}{Fit yield} \\
\hline
\BuToKKK	& 69\,310 & 280 \\
\BuToKpiK	& 5\,760  & 140 \\
\BuToKpipi	& 94\,950 & 430 \\
\BuTopipipi	& 25\,480 & 200 \\
\hline
\hline
\end{tabular}
\end{table}

\section{Systematic uncertainties}
\label{sec:Systematics}

Systematic uncertainties are minimised by measuring the ratios of the $\Bp \to h^+h^{\prime +}h^{\prime -}$ branching fractions relative to one another, but given the statistical precision of the results several sources of significant uncertainty remain.
These originate from possible imperfections in the fit model used to determine the signal yields and the precision with which the relative efficiencies are known.
A summary of the uncertainties assigned to each ratio of branching fractions is given in Table~\ref{table:systematics}.

Pseudoexperiments are used to determine the effect on the signal yields of using alternative shapes to describe the different fit components.
Three variants of the fit model are constructed where in each an alternative shape is used for a particular category of fit component.
In Model~I, the signal and cross-feed components are changed to double Hypatia functions~\cite{Santos:2013gra}.
In Model~II, a set of Chebyshev polynomials up to second order is used to describe the combinatorial background shape.
In Model~III, the partially reconstructed background shapes are replaced with non-parametric functions.
The pseudoexperiments are generated according to the alternative model, then fitted with both the baseline and alternative model.
The mean of the distribution of the difference between the results with the two models is taken as the corresponding systematic uncertainty.
Overall, the Model~II and~III uncertainties are the dominant sources of systematic uncertainty for all measured branching fraction ratios.
Uncertainty from possible bias on the fitted yields is also investigated using pseudoexperiments, generated and fitted using the nominal fit model.
The effect of the fixed parameters in the fit model is estimated by evaluating the impact of varying these parameters within their uncertainties.

Uncertainties on the signal efficiencies originate from residual differences in the behaviour of data and simulation, as well as the limited size of the simulation and control samples.
Data-driven corrections are applied in the determination of the signal efficiency related to the performance of the hardware trigger (denoted L0 \texttt{TOS} and L0 \texttt{TIS} in Table~\ref{table:systematics} for cases where the trigger is associated to the tracks that comprise the \B\ candidate and to other particles in the event, respectively), the reconstruction of tracks, and the \B-meson production kinematics.
The L0 \texttt{TOS} uncertainty is determined from the difference between results with and without the correction applied; this is a more conservative approach compared to those used for other uncertainties, reflecting the fact that the method used to obtain the correction does not account for all possible variables that the efficiency may depend upon.
Effects associated with the reweighting of L0 \texttt{TIS} categories, and with the correction to the track reconstruction efficiency, are both determined by varying the correction within its uncertainties.
The systematic uncertainty associated with the production kinematics correction is estimated by determining the correction factors from an alternative background-subtracted data sample.

Possible small differences between data and simulation in the distribution of the variables included in the MVA are accounted for by weighting the simulated events to match the distributions observed in data.
The changes in results when this weighting is applied are assigned as the associated systematic uncertainties.
Uncertainty in the efficiency of the charm vetoes is obtained by propagating the corresponding values, given in Table~\ref{tab:vetoEfficiencies}.
Effects related to the choice of binning of the efficiency maps are estimated by changing the granularity, while those due to the finite size of the simulated signal samples (denoted ``MC stats'' in Table~\ref{table:systematics}) are evaluated by varying the efficiency maps according to the uncertainties in each SDP bin.
The determination of the PID efficiency from control samples is also a source of uncertainty.
Effects related to the differing kinematic distributions of tracks in the signal modes and the control samples, to the finite size of the control samples, and to the background-subtraction procedure are determined.

\begin{sidewaystable}[!tb]
\caption{
    Absolute systematic uncertainties on the branching fraction ratios.
    All values are given multiplied by 100.
    Uncertainties are presented for all ratios of one mode to another, even though not all are independent.
    }
\label{table:systematics}
\centering
\renewcommand{\arraystretch}{1.5}
\resizebox{0.99\textwidth}{!}{
\begin{tabular}{lcccccccccccccc}
\hline
\hline
\BF ratio
                & Model I & Model II & Model III & Fit bias & Fixed params
                & L0 \texttt{TOS} & L0 \texttt{TIS} & Tracking & Kinematics & MVA
                & Veto & Binning & MC stats & PID \\

\hline
\hline
$\frac{\BF(\BuToKpiK)}{\BF(\BuToKKK)}$
                & \round{2}{0.036} & \round{2}{0.55}  & \round{2}{0.50}  & \round{2}{0.013}  & \round{2}{0.11}
                & \round{2}{0.20}  & \round{2}{0.12}  & \round{2}{0.007}  & \round{2}{0.007} & \round{2}{0.033}
                & \round{2}{0.046}  & \round{2}{0.047}  & \round{2}{0.032}  & \round{2}{0.081}  \\
$\frac{\BF(\BuToKpipi)}{\BF(\BuToKKK)}$
                & \round{1}{0.063} & \round{1}{1. }  & \round{1}{1.2}  & \round{1}{0.038} & \round{1}{0.70}
                & \round{1}{0.79}  & \round{1}{0.38}  & \round{1}{0.20}  & \round{1}{0.24}  & \round{1}{0.61}
                & \round{1}{0.51}  & \round{1}{0.10}  & \round{1}{0.33}  & \round{1}{0.35}  \\
$\frac{\BF(\BuTopipipi)}{\BF(\BuToKKK)}$
                & \round{2}{0.049} & \round{2}{0.017} & \round{2}{0.72}  & \round{2}{0.018} & \round{2}{0.24}
                & \round{2}{0.23}  & \round{2}{0.19}  & \round{2}{0.13}  & \round{2}{0.10}  & \round{2}{0.16}
                & \round{2}{0.12}  & \round{2}{0.36}  & \round{2}{0.11}  & \round{2}{0.16}  \\
\hline
$\frac{\BF(\BuToKKK)}{\BF(\BuToKpiK)}$
                & \round{0}{1.5} & \round{0}{24}  & \round{0}{19}  & \round{0}{0.54}  & \round{0}{4.7}
                & \round{0}{8.5}  & \round{0}{4.9}  & \round{0}{0.28}  & \round{0}{0.30}  & \round{0}{1.4}
                & \round{0}{2.0}  & \round{0}{2.0}  & \round{0}{1.4}  & \round{0}{3.5}  \\
$\frac{\BF(\BuToKpipi)}{\BF(\BuToKpiK)}$
                & \round{0}{2.2}  & \round{0}{32}  & \round{0}{40}   & \round{0}{0.92}  & \round{0}{9.6}
                & \round{0}{9.2}  & \round{0}{8.6}  & \round{0}{0.84}  & \round{0}{1.0}  & \round{0}{1.6}
                & \round{0}{3.3}  & \round{0}{2.7}  & \round{0}{2.4}  & \round{0}{5.5}  \\
$\frac{\BF(\BuTopipipi)}{\BF(\BuToKpiK)}$
                & \round{0}{1.1}  & \round{0}{12}  & \round{0}{14}   & \round{0}{0.28} & \round{0}{2.9}
                & \round{0}{2.6}  & \round{0}{2.6} & \round{0}{0.70} & \round{0}{0.49} & \round{0}{0.37}
                & \round{0}{0.76} & \round{0}{1.4} & \round{0}{0.76} & \round{0}{1.6}  \\
\hline
$\frac{\BF(\BuToKKK)}{\BF(\BuToKpipi)}$
                & \round{1}{0.022} & \round{1}{0.44}  & \round{1}{0.41}  & \round{1}{0.013} & \round{1}{0.24}
                & \round{1}{0.27}  & \round{1}{0.13}  & \round{1}{0.069} & \round{1}{0.082} & \round{1}{0.21}
                & \round{1}{0.17}   & \round{1}{0.036} & \round{1}{0.11}  & \round{1}{0.12}  \\
$\frac{\BF(\BuToKpiK)}{\BF(\BuToKpipi)}$
                & \round{2}{0.040} & \round{2}{0.23}  & \round{2}{0.23}  & \round{2}{0.011} & \round{2}{0.17}
                & \round{2}{0.074} & \round{2}{0.070} & \round{2}{0.007} & \round{2}{0.008} & \round{2}{0.013}
                & \round{2}{0.026} & \round{2}{0.022} & \round{2}{0.019} & \round{2}{0.045}  \\
$\frac{\BF(\BuTopipipi)}{\BF(\BuToKpipi)}$
                & \round{2}{0.040} & \round{2}{0.23}  & \round{2}{0.23}  & \round{2}{0.011} & \round{2}{0.17}
                & \round{2}{0.005} & \round{2}{0.10}  & \round{2}{0.041} & \round{2}{0.017} & \round{2}{0.0082}
                & \round{2}{0.064} & \round{2}{0.20}  & \round{2}{0.062} & \round{2}{0.038}  \\
\hline
$\frac{\BF(\BuToKKK)}{\BF(\BuTopipipi)}$
                & \round{1}{0.21} & \round{1}{0.072} & \round{1}{3.1}  & \round{1}{0.076} & \round{1}{0.94}
                & \round{1}{0.98} & \round{1}{0.77}  & \round{1}{0.53} & \round{1}{0.41}  & \round{1}{0.66}
                & \round{1}{0.50} & \round{1}{1.5}   & \round{1}{0.47} & \round{1}{0.64} \\
$\frac{\BF(\BuToKpiK)}{\BF(\BuTopipipi)}$
                & \round{1}{0.11}  & \round{1}{1.1}  & \round{1}{1.5}   & \round{1}{0.028} & \round{1}{0.26}
                & \round{1}{0.25}  & \round{1}{0.25} & \round{1}{0.068} & \round{1}{0.048} & \round{1}{0.035}
                & \round{1}{0.073} & \round{1}{0.14} & \round{1}{0.071} & \round{1}{0.15} \\
$\frac{\BF(\BuToKpipi)}{\BF(\BuTopipipi)}$
                & \round{1}{0.48}  & \round{1}{2.7} & \round{1}{2.8}  & \round{1}{0.14}  & \round{1}{2.0}
                & \round{1}{0.055} & \round{1}{1.2} & \round{1}{0.50} & \round{1}{0.21}  & \round{1}{0.096}
                & \round{1}{0.78}  & \round{1}{2.4} & \round{1}{0.74} & \round{1}{0.46}  \\
\hline
\hline
\end{tabular}
}
\renewcommand{\arraystretch}{1}
\end{sidewaystable}
\afterpage{\clearpage}

The stability of the results is cross-checked by determining the relative branching fraction ratios in various subsets of the data.
The data are subdivided by year of data-taking and (separately) by magnet polarity, with consistent results obtained.
When comparing results obtained in subsamples separated by hardware trigger decision, by \B-meson pseudorapidity and by detector occupancy some discrepancies can be seen if considering statistical uncertainties alone.
These, however, are compatible with the size of relevant systematic uncertainties.

\section{Results}
\label{sec:Results}

The relative branching fractions of the signal modes are determined, for example with $\BuToKKK$ as denominator, as
\begin{equation}
\frac{\BF(\BuTohhh)}{\BF(\BuToKKK)} = \frac{{\cal N}^{\rm corr}_{hh^{\prime}h^{\prime}}}{{\cal N}^{\rm corr}_{\kaon\kaon\kaon}}\,,
\label{eq:bfeq}
\end{equation}
where ${\cal N}^{\rm corr}$ is, for the mode indicated in the subscript, the efficiency-corrected signal yield accounting both for the variation of the total efficiency across the SDP and for the charm vetoes that completely remove certain regions of the phase space.
These efficiency-corrected yields are~\cite{Pivk:2004ty}
\begin{equation}
{\cal N}^{\rm corr} = \frac{1}{\epsilon^{\rm veto}} \sum_j^{N_{\rm bins}} \frac{ cM_j + \sum_{i \subset {\rm bin}\;j} w_i }{\epsilon^{\rm tot}_j},
\label{eq:Ncorr}
\end{equation}
where the index $j$ runs over the $N_{\rm bins}$ bins of the SDP, $\epsilon^{\rm tot}_j$ is the corresponding efficiency in bin $j$ (as given in Fig.~\ref{fig:totaleffs}), and for each value of $j$ the index $i$ runs over the candidates in that bin.
The per-candidate signal \sWeights\ $w_i$, which implement the background subtraction, are obtained from individual fits to the \B-candidate mass distribution of each mode in which all nuisance parameters are fixed to the values obtained in the simultaneous fit.
In these fits the only varying parameters are the yields of the signal and all background components except those of the cross-feed background contributions, which are fixed.
The term $cM_j$ accounts for these fixed components, where the coefficient $c$ is determined from the fit~\cite{Pivk:2004ty} and $M_j$ is the fraction of the cross-feed background in SDP bin $j$.
The statistical uncertainty on each ${\cal N}^{\rm corr}$ value is calculated as described in Ref.~\cite{LHCb-PAPER-2012-018}, accounting for the reduction in the uncertainties of the yields, compared to the baseline fit, due to the nuisance parameters being fixed.

\begin{table}[!tb]
\caption{
    Measured relative branching fractions of $\Bp \to h^+h^{\prime +}h^{\prime -}$ decays, where the first uncertainty is statistical and the second is systematic.
    Results are presented for all ratios of one mode to another, even though not all are independent.
}
\label{table:final_results}
\centering
\renewcommand{\arraystretch}{1.5}
\begin{tabular}{ll}
\hline
\hline
\BF ratio & Value \\
\hline
    $\BF(\BuToKpiK) / \BF(\BuToKKK)$    & $0.151 \pm 0.004 \pm 0.008$ \\
    $\BF(\BuToKpipi) / \BF(\BuToKKK)$   & $1.703 \pm 0.011 \pm 0.022$ \\
    $\BF(\BuTopipipi) / \BF(\BuToKKK)$  & $0.488 \pm 0.005 \pm 0.009$

    \vspace{1ex}\\\hline

    $\BF(\BuToKKK) / \BF(\BuToKpiK)$    & $6.61\phntm \pm 0.17 \pm 0.33$ \\
    $\BF(\BuToKpipi) / \BF(\BuToKpiK)$  & $11.27      \pm 0.29 \pm 0.54$ \\
    $\BF(\BuTopipipi) / \BF(\BuToKpiK)$ & $3.23\phntm \pm 0.09 \pm 0.19$

    \vspace{1ex}\\\hline

    $\BF(\BuToKKK) / \BF(\BuToKpipi)$    & $0.587\phntm \pm 0.004\phntm \pm 0.008$  \\
    $\BF(\BuToKpiK) / \BF(\BuToKpipi)$   & $0.0888      \pm 0.0023      \pm 0.0047$ \\
    $\BF(\BuTopipipi) / \BF(\BuToKpipi)$ & $0.2867      \pm 0.0029      \pm 0.0045$

    \vspace{1ex}\\\hline

    $\BF(\BuToKKK) / \BF(\BuTopipipi)$      & $2.048 \pm 0.020	\pm 0.040$ \\
    $\BF(\BuToKpiK) / \BF(\BuTopipipi)$     & $0.310 \pm 0.008 \pm 0.020$ \\
    $\BF(\BuToKpipi) / \BF(\BuTopipipi)$    & $3.488 \pm 0.035 \pm 0.053$ \\
\hline
\hline
\end{tabular}
\renewcommand{\arraystretch}{1}
\end{table}

The complete set of results for twelve relative branching fractions of $\Bp \to h^+h^{\prime +}h^{\prime -}$ decays is shown in Table~\ref{table:final_results}.
Six of these are the inverse of the other six.
Moreover, since there are only three independent measurements, correlations between the ratios must also be taken into account.
The statistical and systematic correlations are presented in Tables~\ref{tab:stat-corr} and~\ref{tab:syst-corr}, respectively.
The statistical correlations are determined from ensembles of pseudoexperiments.
In each experiment, the ratios are calculated and the correlation is obtained from the distribution of one ratio against another in the ensemble.
Large statistical correlations are observed between the two ratios that share a decay with a yield that is small compared to that of the other decay channel in the ratios; this affects in particular pairs of ratios that have \BuToKpiK\ as a common channel.
Ratios which do not have any mode in common have smaller correlations, which can however be non-zero due to the nature of the simultaneous fit from which the yields are obtained.

\begin{table}[!tb]
\caption{Statistical correlations between the measured branching fraction ratios.}
\label{tab:stat-corr}
\centering
\renewcommand{\arraystretch}{1.5}
\begin{tabular}{l|cccccc}
\hline\hline
    & $\frac{\pip\Kp\Km}{\Kp\Kp\Km}$ & $\frac{\Kp\pip\pim}{\Kp\Kp\Km}$ & $\frac{\pip\pip\pim}{\Kp\Kp\Km}$ & $\frac{\Kp\pip\pim}{\pip\Kp\Km}$ & $\frac{\pip\pip\pim}{\pip\Kp\Km}$ & $\frac{\pip\pip\pim}{\Kp\pip\pim}$ \\ [0.3ex]
\hline
    $\frac{\BF(\BuToKpiK)}{\BF(\BuToKKK)}$ & --- &\phntn0.16 &\phntn0.10 & $-$0.96 & $-$0.92 & $-$0.01 \\ [0.3ex]
    $\frac{\BF(\BuToKpipi)}{\BF(\BuToKKK)}$ &\phntn0.16 & --- &\phntn0.32 &\phntn0.12 & $-$0.03 & $-$0.34 \\ [0.3ex]
    $\frac{\BF(\BuTopipipi)}{\BF(\BuToKKK)}$ &\phntn0.10 &\phntn0.32 & --- & $-$0.01 &\phntn0.31 &\phntn0.78 \\ [0.3ex]
    $\frac{\BF(\BuToKpipi)}{\BF(\BuToKpiK)}$ & $-$0.96 &\phntn0.12 & $-$0.01 & --- &\phntn0.92 & $-$0.08 \\ [0.3ex]
    $\frac{\BF(\BuTopipipi)}{\BF(\BuToKpiK)}$ & $-$0.92 & $-$0.03 &\phntn0.31 &\phntn0.92 & --- &\phntn0.32 \\ [0.3ex]
    $\frac{\BF(\BuTopipipi)}{\BF(\BuToKpipi)}$ & $-$0.01 & $-$0.34 &\phntn0.78 & $-$0.08 &\phntn0.32 & --- \\
\hline\hline
\end{tabular}
\renewcommand{\arraystretch}{1}
\end{table}

Correlations related to systematic uncertainties obtained from ensembles of pseudoexperiments, as described in Sec.~\ref{sec:Systematics} are evaluated with the same method as the statistical correlations.
For those that are determined from the difference between the results obtained when a single variation is made and those in the baseline analysis, 100\% correlation or anticorrelation (depending on the relative sign of the shift) is assumed.
For each source of systematic uncertainty, these correlations are converted into a covariance matrix.
These are summed, and the total systematic covariance matrix thus obtained is converted back into the total systematic correlation matrix.
The size of the systematic correlations is related to whether two ratios share dominant sources of systematic uncertainty.
In particular, for pairs of ratios with \BuToKpiK\ as a common channel, the uncertainty due to limited knowledge of the background shapes induces significant correlations.

\begin{table}[!tb]
\caption{Systematic correlations between the measured branching fraction ratios.}
\label{tab:syst-corr}
\centering
\renewcommand{\arraystretch}{1.5}
\begin{tabular}{l|cccccc}
\hline\hline
    & $\frac{\pip\Kp\Km}{\Kp\Kp\Km}$ & $\frac{\Kp\pip\pim}{\Kp\Kp\Km}$ & $\frac{\pip\pip\pim}{\Kp\Kp\Km}$ & $\frac{\Kp\pip\pim}{\pip\Kp\Km}$ & $\frac{\pip\pip\pim}{\pip\Kp\Km}$ & $\frac{\pip\pip\pim}{\Kp\pip\pim}$ \\ [0.3ex]
\hline
    $\frac{\BF(\BuToKpiK)}{\BF(\BuToKKK)}$ & --- & $-$0.27 &\phntn0.15 & $-$0.96 & $-$0.97 &\phntn0.38 \\ [0.3ex]
    $\frac{\BF(\BuToKpipi)}{\BF(\BuToKKK)}$ & $-$0.27 & --- &\phntn0.34 &\phntn0.53 &\phntn0.35 & $-$0.72 \\ [0.3ex]
    $\frac{\BF(\BuTopipipi)}{\BF(\BuToKKK)}$ &\phntn0.15 &\phntn0.34 & --- & $-$0.02 &\phntn0.10 &\phntn0.38 \\ [0.3ex]
    $\frac{\BF(\BuToKpipi)}{\BF(\BuToKpiK)}$ & $-$0.96 &\phntn0.53 & $-$0.02 & --- &\phntn0.96 & $-$0.54 \\ [0.3ex]
    $\frac{\BF(\BuTopipipi)}{\BF(\BuToKpiK)}$ & $-$0.97 &\phntn0.35 &\phntn0.10 &\phntn0.96 & --- & $-$0.27 \\ [0.3ex]
    $\frac{\BF(\BuTopipipi)}{\BF(\BuToKpipi)}$ &\phntn0.38 & $-$0.72 &\phntn0.38 & $-$0.54 & $-$0.27 & --- \\
\hline\hline
\end{tabular}
\renewcommand{\arraystretch}{1}
\end{table}

\section{Summary}
\label{sec:Summary}

Data collected by the \lhcb experiment in 2011 and 2012, corresponding to an integrated luminosity of $3.0\invfb$, has been used to determine the relative branching fractions of the $\Bp \to h^+h^{\prime +}h^{\prime -}$ decays.
The measured ratios relative to the \BuToKKK channel are
\begin{align*}
\label{eq:final_results}
\BF(\BuToKpiK) / \BF(\BuToKKK)   &= 0.151 \pm 0.004\stat \pm 0.008\syst \,,\\ \vspace{2ex}
\BF(\BuToKpipi) / \BF(\BuToKKK)  &= 1.703 \pm 0.011\stat \pm 0.022\syst \,,\\ \vspace{2ex}
\BF(\BuTopipipi) / \BF(\BuToKKK) &= 0.488 \pm 0.005\stat \pm 0.009\syst \,.
\end{align*}
The dominant systematic uncertainties are related to knowledge of the background shapes in the invariant-mass fit, and are reducible if knowledge of the various sources of background can be improved or if the background can be suppressed in future analyses.
Several other sources of systematic uncertainty are, however, not negligible compared to the statistical uncertainty of these results, so that further significant reduction in uncertainty will be challenging.

\begin{figure}[!tb]
  \centering
  \includegraphics[width=0.45\textwidth]{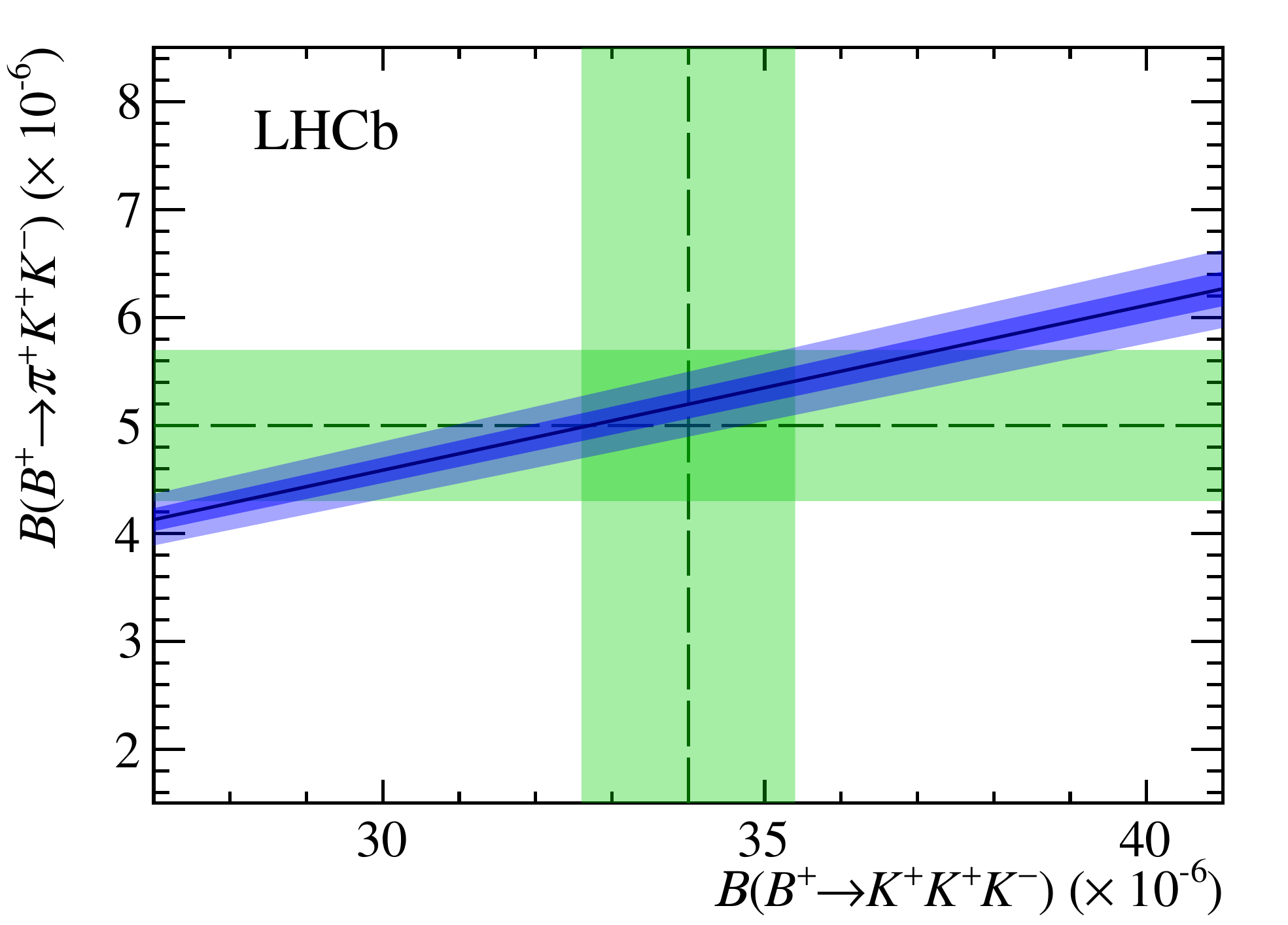}
  \includegraphics[width=0.45\textwidth]{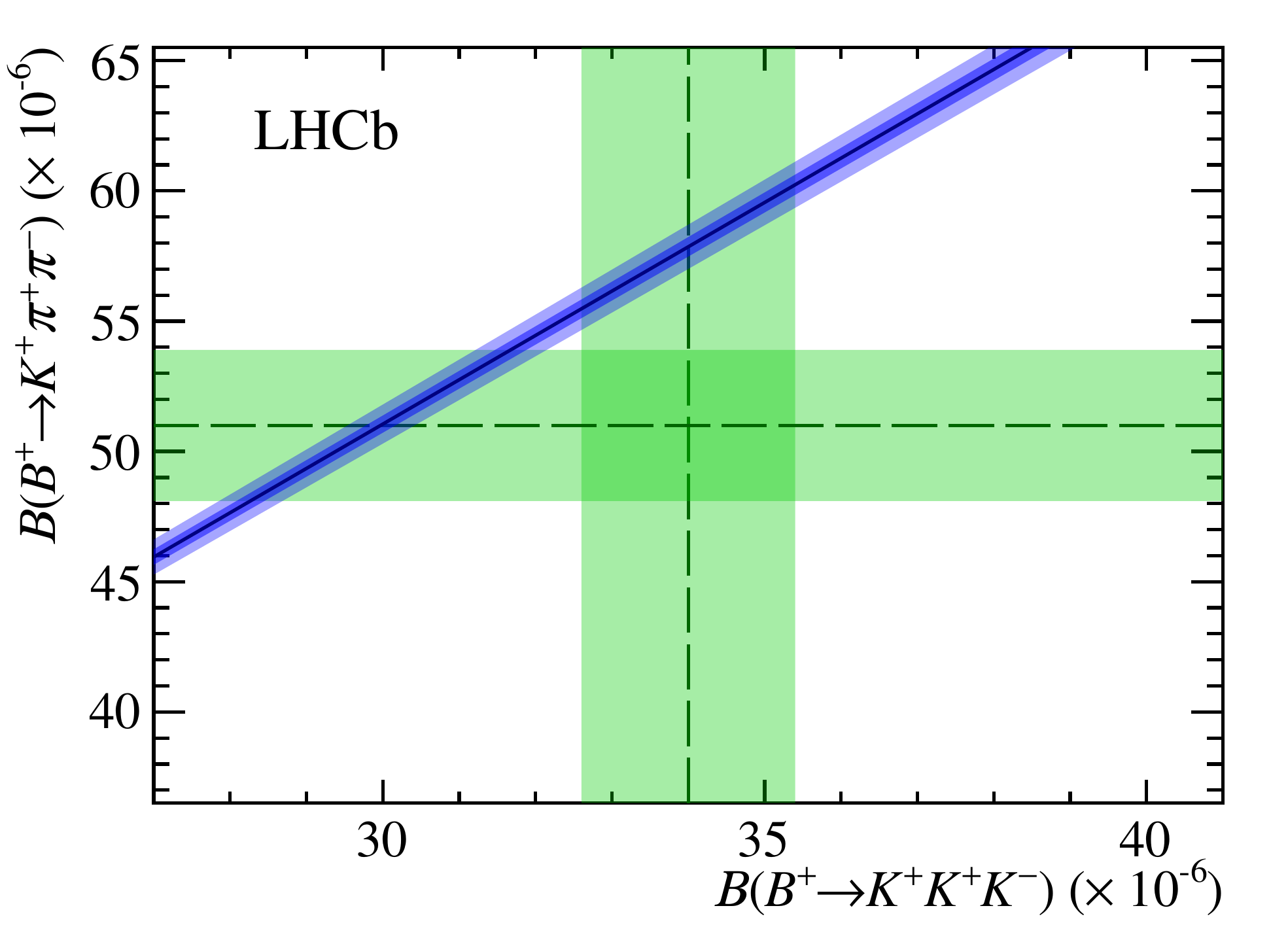}
  \includegraphics[width=0.45\textwidth]{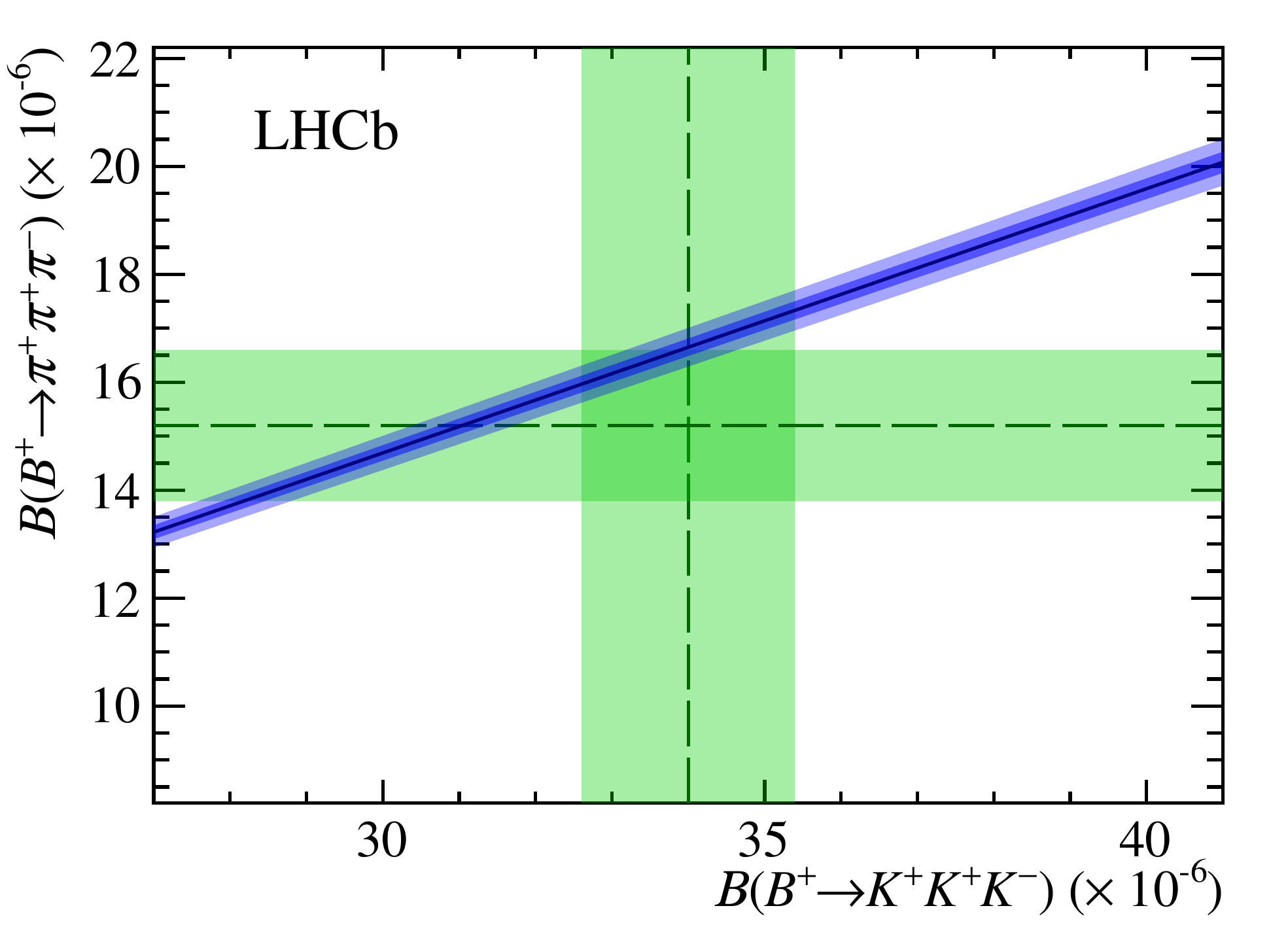}
  \caption{
    Comparisons of the measured branching fraction ratios, with $\BF(\BuToKKK)$ as denominator, with the current world averages~\cite{PDG2020}.
    Light (dark) bands associated with the branching fraction ratio correspond to the $\pm1\sigma$ total (statistical) uncertainty intervals.
    For horizontal and vertical bands taken from the PDG only the total uncertainty is shown.
  }
\label{fig:ratio_limits}
\end{figure}

Comparisons with the current world averages are given, for the three measurements above, in Fig.~\ref{fig:ratio_limits}.
All measurements are in good agreement with the previous world-average results and, furthermore, significant improvement in the precision of all measured ratios is obtained.

\section*{Acknowledgements}

\noindent We express our gratitude to our colleagues in the CERN
accelerator departments for the excellent performance of the LHC. We
thank the technical and administrative staff at the LHCb
institutes.
We acknowledge support from CERN and from the national agencies:
CAPES, CNPq, FAPERJ and FINEP (Brazil); 
MOST and NSFC (China); 
CNRS/IN2P3 (France); 
BMBF, DFG and MPG (Germany); 
INFN (Italy); 
NWO (Netherlands); 
MNiSW and NCN (Poland); 
MEN/IFA (Romania); 
MSHE (Russia); 
MICINN (Spain); 
SNSF and SER (Switzerland); 
NASU (Ukraine); 
STFC (United Kingdom); 
DOE NP and NSF (USA).
We acknowledge the computing resources that are provided by CERN, IN2P3
(France), KIT and DESY (Germany), INFN (Italy), SURF (Netherlands),
PIC (Spain), GridPP (United Kingdom), RRCKI and Yandex
LLC (Russia), CSCS (Switzerland), IFIN-HH (Romania), CBPF (Brazil),
PL-GRID (Poland) and OSC (USA).
We are indebted to the communities behind the multiple open-source
software packages on which we depend.
Individual groups or members have received support from
AvH Foundation (Germany);
EPLANET, Marie Sk\l{}odowska-Curie Actions and ERC (European Union);
A*MIDEX, ANR, Labex P2IO and OCEVU, and R\'{e}gion Auvergne-Rh\^{o}ne-Alpes (France);
Key Research Program of Frontier Sciences of CAS, CAS PIFI,
Thousand Talents Program, and Sci. \& Tech. Program of Guangzhou (China);
RFBR, RSF and Yandex LLC (Russia);
GVA, XuntaGal and GENCAT (Spain);
the Royal Society
and the Leverhulme Trust (United Kingdom).

\addcontentsline{toc}{section}{References}
\bibliographystyle{LHCb}
\bibliography{main,standard,LHCb-PAPER,LHCb-CONF,LHCb-DP,LHCb-TDR}

\newpage
\centerline
{\large\bf LHCb collaboration}
\begin
{flushleft}
\small
R.~Aaij$^{31}$,
C.~Abell{\'a}n~Beteta$^{49}$,
T.~Ackernley$^{59}$,
B.~Adeva$^{45}$,
M.~Adinolfi$^{53}$,
H.~Afsharnia$^{9}$,
C.A.~Aidala$^{84}$,
S.~Aiola$^{25}$,
Z.~Ajaltouni$^{9}$,
S.~Akar$^{64}$
% ...

{\footnotesize \it

% ...
$ ^{9}$Universit{\'e} Clermont Auvergne, CNRS/IN2P3, LPC, Clermont-Ferrand, France\\
% ...
$ ^{25}$INFN Sezione di Milano, Milano, Italy\\
% ...
$ ^{31}$Nikhef National Institute for Subatomic Physics, Amsterdam, Netherlands\\
% ...
$ ^{45}$Instituto Galego de F{\'\i}sica de Altas Enerx{\'\i}as (IGFAE), Universidade de Santiago de Compostela, Santiago de Compostela, Spain\\
% ...
$ ^{49}$Physik-Institut, Universit{\"a}t Z{\"u}rich, Z{\"u}rich, Switzerland\\
% ...
$ ^{53}$H.H. Wills Physics Laboratory, University of Bristol, Bristol, United Kingdom\\
% ...
$ ^{59}$Oliver Lodge Laboratory, University of Liverpool, Liverpool, United Kingdom\\
% ...
$ ^{64}$University of Cincinnati, Cincinnati, OH, United States\\
% ...
$ ^{67}$Syracuse University, Syracuse, NY, United States\\
% ...
$ ^{84}$University of Michigan, Ann Arbor, United States, associated to $^{67}$\\
\bigskip
% ...
\medskip
}
\end{flushleft}

\end{document}